\documentclass[aps,preprintnumbers,amsmath,amssymb,endfloats]{revtex4} % added by PXP Help
\usepackage{dcolumn}% Align table columns on decimal point
\usepackage{bm}% bold math

\pdfoutput=1% for PDFLatex file

   \ifx\pdfoutput\undefined \usepackage[dvips]{graphicx} \else
   \usepackage[pdftex]{graphicx} \pdfcompresslevel=9 \fi
   \usepackage{epstopdf}
   \DeclareGraphicsExtensions{.pdf,.gif,.jpg,.eps,.png}
\begin{document}

\preprint{{\it submitted to The Physics of Plasmas}}

\title{The effects of strong temperature anisotropy on the kinetic
structure of collisionless slow shocks and reconnection exhausts. Part
II: Theory}
 
\author{Yi-Hsin~Liu}
\affiliation{University of Maryland, College Park, MD 20742}
\author{J.~F.~Drake}
\affiliation{University of Maryland, College Park, MD 20742}	
\author{M.~Swisdak}
\affiliation{University of Maryland, College Park, MD 20742}

\date{\today}

\begin{abstract}
Simulations of collisionless oblique propagating slow shocks have
revealed the existence of a transition associated with a critical
temperature anisotropy $\varepsilon=1-\mu_0(P_\|-P_\perp)/ B^2$= 0.25
(Liu, Drake and Swisdak (2011)\cite{yhliu11a}). An explanation for this phenomenon is
proposed here based on anisotropic fluid theory, in particular the
Anisotropic Derivative Nonlinear-Schr\"odinger-Burgers equation, with
an intuitive model of the energy closure for the downstream
counter-streaming ions. The anisotropy value of 0.25 is significant
because it is closely related to the degeneracy point of the slow and
intermediate modes, and corresponds to the lower bound of the coplanar
to non-coplanar transition that occurs inside a compound slow shock
(SS)/rotational discontinuity (RD) wave. This work implies that it is
a pair of compound SS/RD waves that bound the outflows in magnetic
reconnection, instead of a pair of switch-off slow shocks as in
Petschek's model. This fact might explain the rareness of in-situ
observations of Petschek-reconnection-associated switch-off slow
shocks.
\end{abstract}

\maketitle

\section{Introduction}\

Shocks in isotropic MHD have been intensively studied, including the
existence of intermediate shocks (IS) (\cite{brio88a, wu87a,
kennel90a,wu03a} and references therein), the occurrence of dispersive
wavetrains \cite{coroniti71a, hau90a, wu92a}, and the nested subshocks
inside shocks predicted by the Rankine-Hugoniot jump conditions
\cite{kennel88a,longcope10a}. In a collisionless plasma, the effects
of temperature anisotropy need to be considered, which can be done for
linear waves with the Chew-Goldberger-Low (CGL) framework
\cite{abraham-shrauner67a, hau93a}. Hau and Sonnerup have pointed out
the abnormal properties of the linear slow mode under the influence of
a firehose-sense ($P_\| > P_\perp$) pressure anisotropy, including a
faster phase speed compared to the intermediate mode, a fast-mode-like
positive correlation between magnetic field and density, and the
steepening of the slow expansion wave. In kinetic theory both
anisotropy and high $\beta$ can greatly alter the linear mode behavior
\cite{krauss-varban94a, karimabadi95a}. The anisotropic
Rankine-Hugoniot jump conditions have been explored while taking the
downstream anisotropy as a free parameter \cite{chao70a, hudson70a,
hudson71a, hudson77a}, while Hudson \cite{hudson71a} calculated the
possible anisotropy jumps across an anisotropic rotational
discontinuity. Karimabadi {\it et al.} \cite{karimabadi95a} noticed
the existence of a slow shock whose upstream and downstream are both
super-intermediate. But, a comprehensive nonlinear theory describing
the coupling between slow and intermediate shocks under the influence
of temperature anisotropy has not yet been presented.

\begin{figure}
\includegraphics[width=13cm]{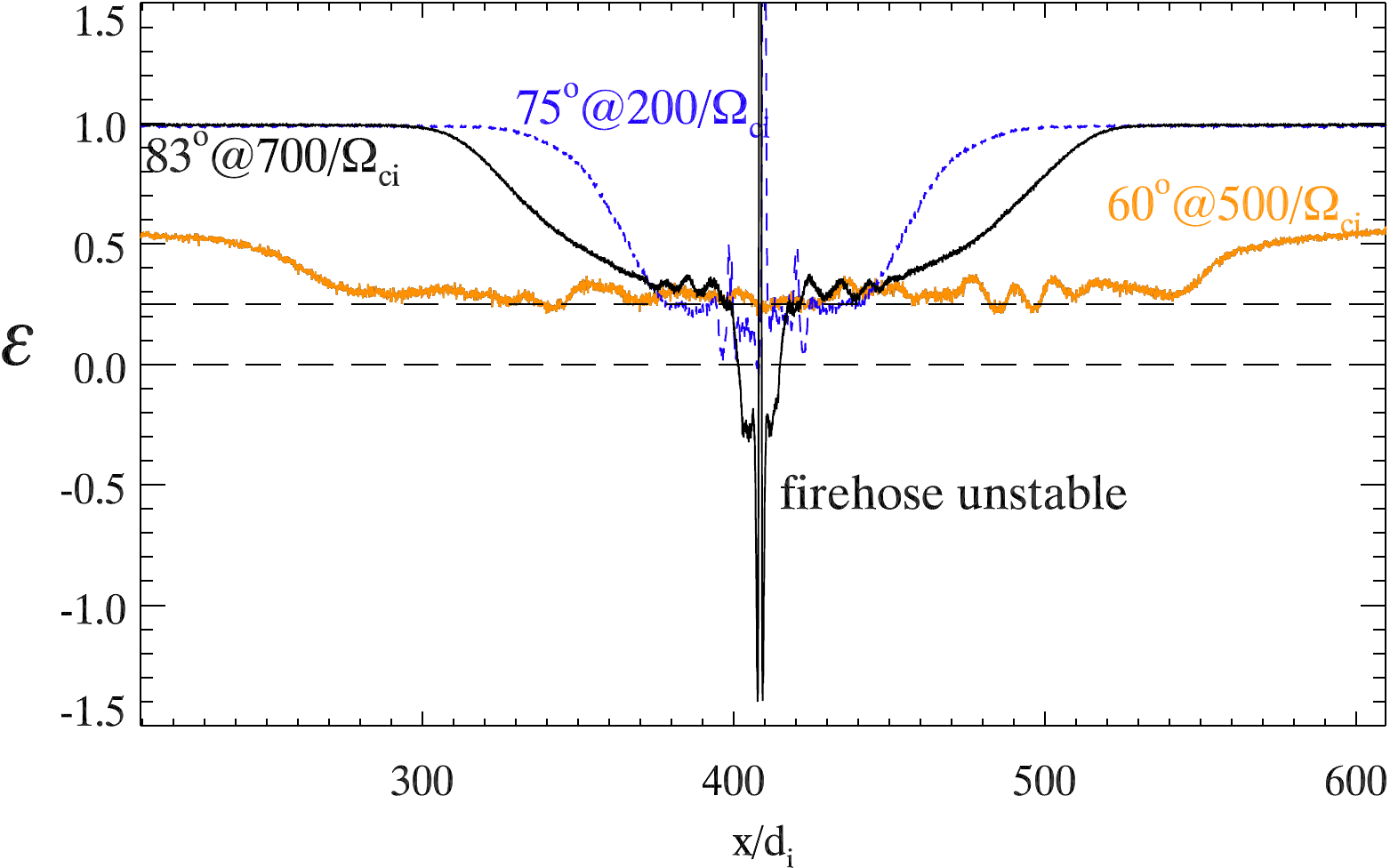} 
\caption{ [From Liu {\it
et al.}, (2011) \cite{yhliu11a}] The $\varepsilon$ distributions of
runs with $\theta_{BN}=60^\circ$ at $500/\Omega_{ci}$, $75^\circ$ at
$200/\Omega_{ci}$, $83^\circ$ at $700/ \Omega_{ci}$. $\theta_{BN}$ is
the angle between the magnetic field and the shock normal direction
($\hat{{\bf e}}_x$) far upstream in these shock
simulations. $\Omega_{ci}$ is the ion cyclotron frequency based on the
upstream magnetic field. These simulations are designed to study the
structure of reconnection exhausts in the normal direction. The
downstream $\varepsilon$ tends to plateau at $0.25$. When $\varepsilon
< 0$, the plasma is susceptible to the firehose instability.}
\label{LO_epsilons} \end{figure}

In Petschek's description of magnetic reconnection, the reconnection
exhaust is bounded by a pair of back-to-back standing switch-off slow
shocks. Particle-in-cell (PIC) simulations of such shocks
\cite{yhliu11a, yin07b} exhibit large downstream
temperature anisotropies.  In Liu {\it et al.} (2011) \cite{yhliu11a}
(hereafter called Paper I) and Fig.~\ref{LO_epsilons} of this paper we
show that when the parameter $\varepsilon=
1-\mu_0(P_{\|}-P_{\bot})/B^2=0.25$, the behavior of the coplanar shock
undergoes a transition to non-coplanar rotation.
This firehose-sense temperature anisotropy slows the linear
intermediate mode and speeds up the linear slow mode enough so that,
at some point, their relative velocities can be reversed
\cite{abraham-shrauner67a, hau93a}. This reversal is reflected in
the structure of the Sagdeev potential (also called the
pseudo-potential) \cite{sagdeev66a}, which characterizes the
nonlinearity of the system. In this work a simplified theoretical
model is developed to explore the effect of temperature anisotropy on
the structure of the Sagdeev potential and to provide an explanation
for the extra transition inside the switch-off slow shock (SSS)
predicted by isotropic MHD.  The theory suggests that in PIC
simulations a compound slow shock (SS)/ rotational discontinuity (RD)
is formed instead of a switch-off slow shock.  This work may help to
explain satellite observations of compound SS/RD waves
\cite{whang96a, whang97a}, anomalous slow shocks
\cite{walthour94a} and the trapping of an RD by the internal
temperature anisotropy of a slow shock in hybrid simulations
\cite{LCLee00a}.

In Sec. II of this paper we introduce our model equations for studying
the nonlinear coupling of slow and intermediate waves under the
influence of a temperature anisotropy. In Sec. III we calculate the
speeds and the eigenmodes of slow and intermediate waves. Sec. IV
points out the existence of extra degeneracy points between slow and
intermediate modes introduced by the temperature anisotropy, and
comments on the consequences (in the context of the Riemann problem)
of having the slow wave faster than the intermediate wave. In Sec. V
we introduce a simple energy closure. In Sec. VI. A. we calculate the
pseudo-potential of stationary solutions, and apply the equal-area
rule to identify the existence of compound SS/RD waves and compound
SS/IS waves. In Sec. VI. B we demonstrate the significance of
$\varepsilon=0.25$ as being the lower bound of the SS to RD transition
in compound SS/RD waves. In Sec. VII we discuss the time-dependent
dynamics that help keep $\varepsilon=0.25$. In Sec VIII we provide more
evidence from PIC simulations to support the existence of compound
SS/RD waves at the boundaries of reconnection exhausts. In Sec IX, we
summarize the results and point out the relation between compound
SS/RD waves and anisotropic rotational discontinuities
\cite{hudson71a}.

\section{The Anisotropic Derivative Nonlinear Schr\"odinger-Burgers equation}\label{seceqns}
Instead of analyzing the anisotropic MHD equations, which have seven
characteristics (waves), we simplify the system into a model equation
that possesses only two characteristics. This model equation will be
ideal for demonstrating the underlying coupling between the nonlinear
slow and intermediate modes. Beginning with the anisotropic MHD
equations \cite{chao70a}, we follow the procedure of Kennel {\it et
al.}
\cite{kennel90a,kennel88b} to derive the Anisotropic
Derivative Nonlinear Schr\"odinger-Burgers equation (ADNLSB) (see
Appendix I for details),
\begin{equation}
\partial_\tau { \bf b}_t + \partial_\eta[\alpha{\bf b}_t
(b_t^2-b_{t0}^2)+\Omega{\bf b}_t(\varepsilon-\varepsilon_0)]
=\partial_\eta(R\partial_\eta {\bf
b}_t)-\frac{1}{2\sqrt{\varepsilon_0}}d_i\partial_\eta^2(\hat{\bf e}_x
\times {\bf b}_t)
\label{ADNLSB}
\end{equation}
This equation describes waves that propagate in the $x$-direction in the
upstream intermediate speed frame. In this frame $\eta\equiv
x'-\sqrt{\varepsilon_0}C_{An}t$ is the spatial coordinate with
$C_{An}^2 \equiv B_x^2/(\mu_0 \rho_0)$, and $\tau\equiv
\sqrt{\varepsilon_0}C_{An}t$ is the time used to measure the slow
variations (such as steepening processes). ${\bf b}_t = {\bf
B}_t/B_x$, where the subscript ``$t$'' represents the component
tangential to the wave-vector and here will be in the $y$-$z$
plane. The anisotropy parameter $\varepsilon
=1-\mu_0(P_{\|}-P_{\bot})/B^2$. The subscript ``0'' denotes the
upstream parameters. The right hand side term proportional to the ion
inertial length, $d_i\equiv \sqrt{m_i/(\mu_0 n e^2)}$, represents
dispersion (which can be viewed as the spreading tendency of Fourier
decomposed waves of different wavenumbers), while the term containing
$R$ describes dissipation from magnetic resistivity. Here $R$ is a
constant. The terms proportional to $\alpha$ and $\Omega$ are the nonlinearities of this wave
equation, where
\begin{equation}
\alpha\equiv\frac{[(3\gamma-1)\varepsilon_0-(3\gamma-4)]C_{An}^2}{12[(\varepsilon_0-A)
C_{An}^2-C_S^2]}
\end{equation}
\begin{equation}
\Omega\equiv\frac{(b_{t0}^2-2)C_{An}^2}{6[(\varepsilon_0-A)C_{An}^2-C_S^2]}+
\frac{1}{2\varepsilon_0}
\end{equation}
and
\begin{equation}
A\equiv\frac{2}{3}(\gamma-1)(1-\varepsilon_0)(1+b_{t0}^2)
\end{equation}
Here $C_S^2 \equiv \gamma P_0/\rho_0$ and $\gamma=5/3$ for monatomic
plasma. Since we are studying reconnection exhausts, $\theta_0$ (the
angle between the upstream magnetic field and $\hat{{\bf e}}_x$) is
typically large ($\sim 80^\circ$). Therefore $\beta_n\equiv
\beta/\cos^2\theta_0=C_S^2/C_{An}^2 \gg 1$, and
$(\varepsilon_0-A)C_{An}^2-C_S^2 \sim -C_S^2 < 0$. Hence this equation
describes only the slow and intermediate modes \cite{kennel90a} (this
is shown explicitly in the next section). This fact relates to the
degeneracy properties of ideal MHD for parallel propagating waves,
namely that the fast and intermediate modes degenerate in $\beta <1$
plasmas, while the slow and intermediate mode degenerate in $\beta>1$
plasmas.  Finally, we note that Eq.~(\ref{ADNLSB}) is applicable in
the weak nonlinearity limit.

\section{The conservative form- wave propagation}\
  
In order to explore the structure of the reconnection exhaust, a
comprehensive understanding of how waves connect to each other across
a transition is required. This is called a Riemann problem. Neglecting
the source terms on the right hand side (RHS), the left hand side
(LHS) of Eq.~(\ref{ADNLSB}) is a hyperbolic equation in conservative
form.
 
Letting $\varepsilon - \varepsilon_0 \equiv \delta
\varepsilon(b_z,b_y) $, and ${\bf b}_{t0}= b_{z0} {\bf \hat{e}}_z$, we
obtain,
\begin{equation}
\partial_\tau {\bf q}+\partial_\eta {\bf f(q)}=0,
\end{equation}
with
\begin{align}
{\bf q}\equiv \left [
     \begin{array}{clr}
     q_1\\
     q_2
     \end{array} \right]
     =\left [
     \begin{array}{clr}
      b_z \\
      b_y 
     \end{array} \right],  \qquad &
{\bf f} \equiv \left [
     \begin{array}{clr}
     f_1\\
     f_2
     \end{array} \right]
       =\left [
     \begin{array}{clr}
     \alpha b_z (b_z^2+b_y^2-b_{z0}^2) + \Omega b_z \delta \varepsilon\\
     \alpha b_y (b_z^2+b_y^2-b_{z0}^2) + \Omega b_y \delta \varepsilon
     \end{array} \right].
     \label{flux}
\end{align} 
We can obtain the characteristics (waves) of this equation by
analyzing its flux function, {\bf f}.  Its Jacobian is
\begin{equation}
\partial_q f= \left [
     \begin{array}{clr} \alpha(3b_z^2+b_y^2-b_{z0}^2)+\Omega(\delta
     \varepsilon+b_z \delta \varepsilon_{b_z}) & 2\alpha b_zb_y
     +\Omega b_z \delta \varepsilon_{b_y}\\ 2\alpha b_zb_y+\Omega b_y
     \delta \varepsilon_{b_z} & \alpha(3b_y^2+b_z^2-b_{z0}^2+\Omega
     (\delta\varepsilon +b_y \delta \varepsilon_{b_y})) \end{array}
     \right],
\end{equation} 
where $\delta \varepsilon_{b_z} \equiv \partial (\delta
\varepsilon)/\partial b_z$, $\delta \varepsilon_{b_y} \equiv \partial
(\delta \varepsilon)/\partial b_y$. One eigenvalue (also called the
characteristic speed) is
\begin{equation}
\lambda_{SL} =\alpha(3b_t^2-b_{t0}^2) + \Omega (\delta \varepsilon+
b_z\delta \varepsilon_{b_z}+ b_y \delta \varepsilon_{b_y})
\label{lambda_SL}
\end{equation} 
with eigenvector
\begin{equation}
{\bf r}_{SL}= \frac{1}{b_t}\left [
     \begin{array}{clr}
     b_z\\
     b_y
     \end{array} \right].
\end{equation} 

In isotropic ideal MHD, where the $(\varepsilon-\varepsilon_0)$ term
is dropped, the eigenvalue in the infinitesimal limit ($b_t
\rightarrow b_{t0}$) is $\lambda_{SL}=2\alpha b_{t0}^2$, which is the
phase speed of the linear slow mode in the intermediate mode
frame. The subscript ``SL'' means the slow mode. The eigenvector
indicates that slow mode is coplanar ({\it i.e.,} in the radial
direction in $b_z-b_y$ space).

The other mode has eigenvalue
\begin{equation}
\lambda_{I} =\alpha(b_t^2-b_{t0}^2) + \Omega \delta \varepsilon
\label{lambda_I_0}
\end{equation} 
and eigenvector
\begin{equation}
{\bf r}_{I} \varpropto \left [
     \begin{array}{clr}
     \alpha b_y +\Omega \delta \varepsilon_{b_y}/2\\
     -\alpha b_z -\Omega \delta \varepsilon_{b_z}/2
     \end{array} \right].
\end{equation} 
In isotropic ideal MHD, where the $(\varepsilon-\varepsilon_0)$ term
is dropped, the eigenvalue in the infinitesimal limit is
$\lambda_{I}=0$, which is the phase speed of the linear intermediate
mode in the intermediate mode frame. The subscript ``I'' means the
intermediate mode. The eigenvector indicates that this intermediate
mode is non-coplanar ({\it i.e.,} in a non-radial direction).
 
It can be shown that $\nabla_{\bf q} \lambda_{I}({\bf q}) \cdot {\bf
r}_{I}({\bf q})=0$ for all ${\bf q}$. This means that along the
eigen-direction of the intermediate mode the characteristic speed is
constant, and thus the mode exhibits no steepening or spreading, just
as is the case for its counterpart in isotropic MHD. (This behavior is
also confirmed by the anisotropic MHD simple wave calculation; see
Appendix II). Therefore the intermediate mode is termed ``linearly
degenerate".  If we are looking for a transition in the $-{\bf
r}_{SL}$ direction (toward $b_t=0)$, then the portion of the slow mode
with $\nabla_{\bf q} \lambda_{SL}({\bf q}) \cdot {\bf r}_{SL}({\bf q})
<0$ will steepen into a slow shock. When $\lambda_{SL}$ at the
downstream of a transition is larger than that at the upstream, the
downstream wave will catch up with the upstream wave and thus steepen.

\section{A new degeneracy point due to the temperature anisotropy}\

\begin{figure}
 \includegraphics[width=13cm]{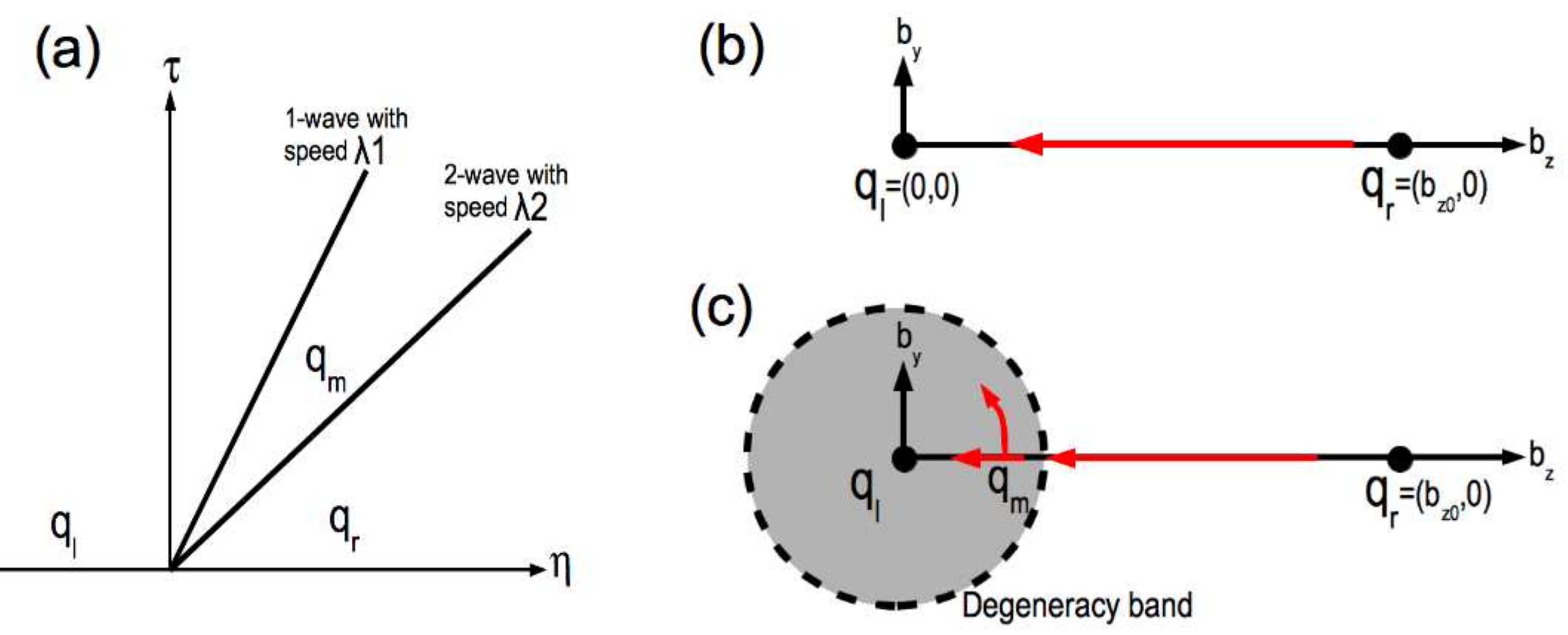} 
\caption{Panel (a): An initial discontinuity between ${\bf q}_r$ and
${\bf q}_l$ results in two waves, the ``1-wave'' and ``2-wave'', that
propagate in the $\eta$ direction along time $\tau$. The middle state
${\bf q}_m$ needs to be determined; Panel (b): The state space plot in
the $(b_z,b_y)$ plane. The value ${\bf q}_r=(b_{z0},0)$ is chosen
since there is no out-of-plane $B_y$ upstream of the slow shocks in
Paper I. ${\bf q}_r$ straightly connects to ${\bf q}_l$ and forms a
switch-off slow shock; Panel (c): In order to connect ${\bf q}_r$ to
${\bf q}_l$, it is necessary to cross the degeneracy band into the
reversal region, which could cause the path to rotate at ${\bf q}_m$.}
\label{degeneracy} \end{figure}

In the Riemann problem for our two mode system, we seek to determine
the middle state ${\bf q}_m$ that connects the faster ``2-wave'' from
a given state ${\bf q}_r$, to the slower ``1-wave'' from a given state
${\bf q}_l$ (see Fig.~\ref{degeneracy} (a); the subscripts ``r'' and
``l'' mean right and left respectively). In order to determine the
path that connects ${\bf q}_r$ to ${\bf q}_l$ in the state space
($b_z-b_y$ space, in this case), the Hugoniot locus that connects
${\bf q}_l$ or ${\bf q}_r$ to a possible asymptotic state by shock
waves needs to be calculated, as do the integral curves for possible
rarefaction waves (see, for example, \cite{leveque02a}). The Hugoniot
locus in state space is a curve formed by allowing one of the
parameters in the standard Rankine-Hugoniot jump condition to
vary. The integral curve is formed by following the eigenvector from a
given state in state space.
 
In order to proceed we further assume a gyrotropic energy closure,
$\delta \varepsilon=\delta \varepsilon(b_t)$, which allows us to write
\begin{align}
{\bf r}_{SL}= \frac{1}{b_t}\left [
     \begin{array}{clr}
     b_z\\
     b_y
     \end{array} \right] &&
     {\bf r}_{I}= \frac{1}{b_t}\left [
     \begin{array}{clr}
     b_y\\
     -b_z
     \end{array} \right] &&
     \lambda_{SL}-\lambda_{I}=2\alpha b_t^2 + \Omega\delta\varepsilon_ {b_t} b_t
\end{align} 
The Hugoniot locus and integral curves of the intermediate mode are
identical in our system. This is also the case for the slow mode (we
perform the calculation in Appendix III). For a given state, ${\bf
q}_r={\bf b}_{t0}=(b_{z0}, 0)$, the Hugoniot locus and integral curve
of the intermediate mode are
\begin{equation}
b_y^2=b_{z0}^2-b_z^2
\end{equation} 
which is a circle in state space. Note that even though we can
calculate the Hugoniot locus and integral curve for the intermediate
mode, the solution is the same as for a finite amplitude intermediate
mode that does not steepen into a shock or spread into a rarefaction.
For the slow mode, the Hugoniot locus and integral curve are
\begin{equation}
b_y=0
\end{equation} 
which is in the radial direction in state space. This direction
implies that the slow shock is coplanar, even in the presence of
temperature anisotropy, just as is the case for its counterpart in
full anisotropic MHD \cite{chao70a}. In the isotropic case, this curve
forms a slow shock if the path is toward the origin, and a slow
rarefaction if the path is away from the origin.

The state ${\bf q}_r$ can connect to ${\bf q}_l=(0,0)$ by following
the Hugoniot locus of a slow mode that starts from ${\bf q}_r$, as
shown in Fig.~\ref{degeneracy}(b).  In isotropic MHD this forms a
switch-off slow shock. However, a strong enough temperature anisotropy
introduces new degeneracy points (which occur where
$\lambda_{SL}-\lambda_I=0$) when $2\alpha b_t +
\Omega\delta\varepsilon_ {b_t}=0$, other than the traditional
degeneracy point at $b_t=0$. These points form a band circling the
origin as shown in Fig.~\ref{degeneracy}(c). Inside the band, the
intermediate mode is slower than the slow mode. Physically, this
implies that a rotational intermediate mode can arise downstream of a
slow mode, something which is not allowed in a Riemann problem in
isotropic MHD. This effect is realized when the path along the
Hugoniot locus ($-{\bf r}_{SL}$ direction) of the slow mode from ${\bf
q}_r$ switches to the solution of the intermediate mode (circular
direction) somewhere (${\bf q}_m$) inside the degeneracy band.

This behavior can explain the morphological differences between the
shock simulations in the cases $\theta_{BN}=30^\circ,45^\circ$ and
those for $\theta_{BN}=60^\circ,75^\circ,83^\circ$ of Paper I
\cite{yhliu11a}. The latter has an extra transition to the rotational
direction that is similar to the path in Fig.~\ref{degeneracy}(c). We
now look for a similar effect in state space and a way of determining
${\bf q}_m$ in a more detailed analytical model.

\section{An energy closure based on counter-streaming ions}\

\begin{figure}
\includegraphics[width=10cm]{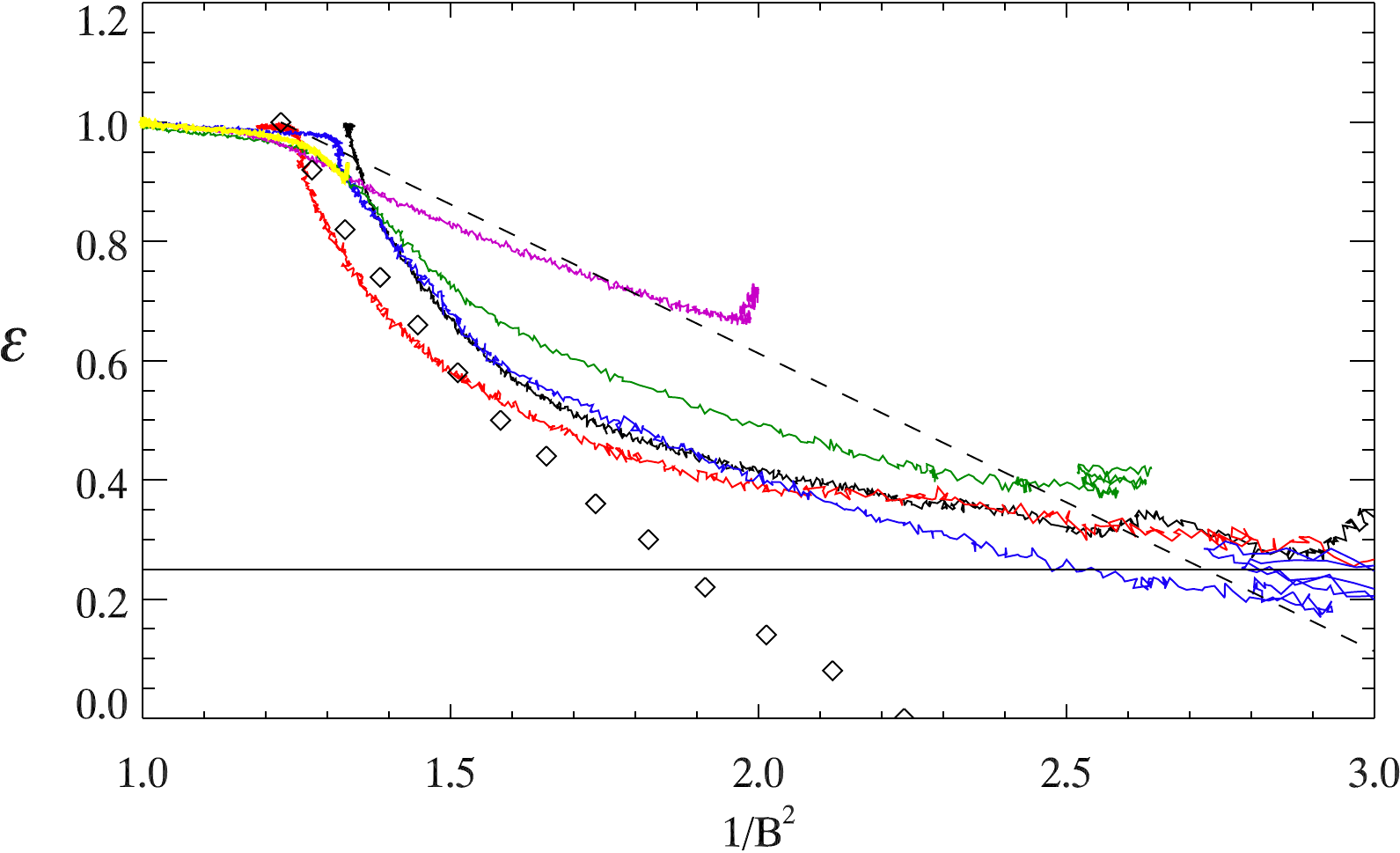} 
 \caption{ The $\varepsilon$ distribution vs. $1/B^2$ for the cases
$\theta_{BN}=30^\circ$(yellow), $45^\circ$(magenta),
$52^\circ$(green), $60^\circ$(blue), $75^\circ$(red) and
$83^\circ$(black) from Fig. 5 of Paper I \cite{yhliu11a}. The dashed
line has slope -0.5. In comparison, the diamond curve is the
theoretical prediction with the CGL condition for the
$\theta_{BN}=75^\circ$ case.}  \label{LO_closure} \end{figure}

In order to close the ADNLSB equations, we need a energy closure
$\varepsilon(b_t)$. The modeling of the energy closure for a
collisionless plasma has historically been difficult. The
Chew-Goldberger-Low (CGL) condition \cite{chew56a} is one choice, but
it does not work well when streaming ions are present. Since we are
here just trying to qualitatively demonstrate the underlying physics,
we will assume that we have a $\varepsilon(b_t)$, where $\varepsilon$
and $|B|$ are simply related by
\begin{equation}
\varepsilon = c_1- \frac{c_2}{B^2}
\label{closure}
\end{equation} 
with positive constants $c_1$ and $c_2$ and the condition
$c_1-c_2/B_0^2=\varepsilon_0$ is imposed. This functional form is
motivated by the nearly constant parallel pressure maintained by
free-streaming ions ($c_2 \sim P_{\|}$). Although Eq.~(\ref{closure})
is strictly empirical, results from PIC simulations (see
Fig.~\ref{LO_closure}) suggest that $c_2= 0.5$ provides a reasonable
first approximation and will be used in the following calculations.

We take the variation,
\begin{equation}
\delta \varepsilon= \varepsilon-\varepsilon_0 \sim \frac{c_2}{B_x^2 b^4}\delta (b_t^2)= \frac{c_2}{B_x^2 b^4}(b_t^2-b_{t0}^2) 
\end{equation} 
where $b^2=1+b_t^2$. This parameterization will be valid whenever
$\delta\varepsilon(b_t) \ll \varepsilon_0$. Therefore, an effective
nonlinearity in Eq.~(\ref{ADNLSB}) can be written as,
\begin{equation}
\alpha_\text{eff}(b_t)=\alpha +  \Omega \frac{c_2}{B_x^2 b^4}
\label{alpha_eff}
\end{equation}  
The most important conclusions in the remainder of this work do not
depend on the details of the closure, but only that $\varepsilon$
decreases as $|B|$ decreases. Note that $\Omega
\sim 1/2\varepsilon_0$ is mostly positive in the limit in which we are
interested. This fact will be used in the following section.

\section{The pseudo-potential:  Looking for a stationary solution}\
\subsection{The formation of compound SS/RD waves and SS/IS waves}  
In order to determine both where the path in the state space of
Fig.~\ref{degeneracy}(c) will turn to the intermediate rotation and
the nontrivial coupling of the slow and intermediate modes when
temperature anisotropies are present, we construct the
pseudo-potential of a stationary solution.  We will look for a
equation that possesses traveling stationary waves, by substituting
${\bf b}_t={\bf b}_t[\xi(\eta -V_S\tau)]$ (where $V_S$ is the speed of
the stationary wave observed in the upstream intermediate frame) into
Eq.~(\ref{ADNLSB}) and integrating over $\xi$ once.  We obtain
\begin{equation}
R \partial_\xi {\bf
b}_t-\frac{1}{2\sqrt{\varepsilon_0}}d_i\partial_\xi(\hat{\bf e}_x
\times {\bf b}_t)=-V_S({\bf b}_t-{\bf b}_{t0}) +
\alpha_\text{eff}(b_t){\bf b}_t (b_t^2-b_{t0}^2)\equiv - {\bf F}
\equiv \partial_{{\bf b}_t} \Psi .
\label{stationary}
\end{equation}
In this formulation we can treat ${\bf b}_t$ as a spatial coordinate,
and $\xi$ as time. The terms on the LHS of Eq.~(\ref{stationary})
behave analogously to, respectively, a frictional force and a Coriolis
force with rotational frequency $d_i/\sqrt{\varepsilon_0}$ and
rotational axis $\hat{\bf e}_x$. A pseudo-potential $\Psi$ that
characterizes the nonlinearity is uniquely defined because $\partial
F_z/\partial b_y=\partial F_y/\partial b_z$. We are only interested in
the small $V_S$ limit, because the upstream values with subscript
``0'' are expected to be the upstream values of a switch-off slow
shock in ideal isotropic MHD, which propagates at the upstream
intermediate speed. The anisotropy in our PIC simulations does not
seem to significantly change this behavior \cite{yhliu11a}.

Calculating the pseudo-work done on the pseudo-particle, $\int
[\text{Eq.}(\ref{stationary})] \cdot \partial_\xi {\bf b}_t d\xi$, we
obtain \begin{equation}
\Psi|_\text{up}^\text{down}=R \int_\text{up}^\text{down} (\partial_\xi {\bf b}_t)^2 d\xi < 0
\end{equation}
Note that from upstream to downstream is in the negative $\xi$
direction.  The pseudo-particle will move to a lower potential, while
its total energy is dissipated by the resistivity and the rate of the
drop depends on the strength of the resistivity. Kennel {\it et al.}
\cite{kennel90a} have shown that when pseudo-particles move toward
lower pseudo-potentials, the entropy increases and so the resulting
shock is admissible. Note that the Coriolis-like force does not do
work. It only drives rotation of the pseudo-particle on the
iso-surface of the pseudo-potential and hence causes stable nodes to
become stable spiral nodes and unstable nodes to become unstable
spiral nodes, thus leading to the formation of dispersive wavetrains
\cite{hau90a, wu92a}. We will neglect its effect in the following
discussion.

The pseudo-potential is shown in Fig.~\ref{potential}(a) for the
parameters $\theta_0=42^\circ$, $\beta_0=1$, $\varepsilon_0=1$,
$c_2=0.5$ and $V_S =0$. The temperature anisotropy has turned the
origin from a local minimum of the pseudo-potential in the isotropic
MHD model to a local maximum. We term this the ``reversal
behavior''. A pseudo-particle initially at point ${\bf q}_r$ will
slide down the hill in the slow mode eigen-direction, and then follow
the circular valley in the intermediate eigen-direction. Without the
reversal behavior (e.g., in isotropic MHD) the pseudo-particle will
slide down to the origin and form a switch-off slow shock. The
trajectory of the pseudo-particle can also be calculated by
numerically integrating Eq.~(\ref{stationary}) with respect to
$\xi$. In (b) the variation of the temperature anisotropy is
shown. Similar reversal behaviors can be found in fully anisotropic
MHD with the energy closure used here, or with the CGL closure (see
Appendix IV.~A).

In Fig.~\ref{potential} (c), we plot a cut of the pseudo-force, $F_z$,
effective $\alpha_{\text{eff}}$ and the pseudo-potential $\Psi$ along
the $b_z$ axis ($b_y=0$), which is the eigen-direction of a slow mode
beginning at ${\bf q}_r=(b_{z0},0)$. Here
\begin{equation}
F_z=-\alpha_\text{{eff}}(b_z) b_z (b_z^2-b_{z0}^2) + V_S (b_z-b_{z0}). 
\label{Fz}
\end{equation}
It is clear that $\Psi_{min}$ occurs at $F_z=0$ in
Fig.~\ref{potential}(c), since $\Psi$ is constructed by integrating
the pseudo-force ${\bf F}$.  In Fig.~\ref{potential} (d) we plot cuts
of the characteristics of the slow and intermediate modes along the
$b_z$ axis.  \begin{equation}
\lambda_{SL}=\alpha (3b_z^2-b_{z0}^2) +\Omega \frac{c_2}{B_x^2} \left[\left(\frac{1}{b^4}-4\frac{b_z^2}{b^6}\right)(b_z^2-b_{z0}^2)+2\frac{b_z^2}{b^4}\right]
\end{equation}
 \begin{equation}
\lambda_{I}=\alpha_\text{{eff}}(b_z) (b_z^2-b_{z0}^2)
\label{lambda_I} 
\end{equation}
The temperature anisotropy has changed the structure of these
characteristics. As a result, there are new degeneracy points
($\lambda_{SL}=\lambda_I$) between slow and intermediate waves such as
the point ``D''. The slow characteristic shows extra nonconvexity
points, where no steepening and spreading occurs ({\it i.e.},
$\nabla_{\bf q} \lambda_{SL}({\bf q}) \cdot {\bf r}_{SL}({\bf q})=0$),
such as the point ${\bf a}$ (the local maximum of
$\lambda_{SL}$). This is clearer when we compare the slow
characteristic here to that in the isotropic case shown in
Fig.~\ref{SS2IS}, where $b_t=0$ is the only degeneracy point and the
nonconvexity point of the slow mode.
%The $\lambda_I$ can always be written in this simple form.

In order to identify the nonlinear waves determined by the route of
the pseudo-particle, we apply the equal-area rule, which tells how
shocks are steepened from characteristics. The equal-area rule (see
Appendix V for more details) applied to $\lambda_{SL}$ shows that the
sliding route (point ${\bf q}_r$ to ${\bf q}_m$) forms a slow
shock. Since $\nabla_{\bf q} \lambda_{SL}({\bf q}) \cdot {\bf
r}_{SL}({\bf q}) <0$ therefore $\lambda_{SL}|_{{\bf q}_m} >
\lambda_{SL}|_{{\bf q}_r} $, and thus the slow mode will steepen until
the red area above the horizontal line $V_S=0$ equals the red area
below $V_S=0$. The slow shock transition immediately connects to the
intermediate mode (point ${\bf q}_m$ to ${\bf b}$, which is also from
the equal-area rule on $\lambda_{SL}$) in the valley. The fact that both
the upstream (point ${\bf q}_m$) and downstream (point ${\bf b}$)
travel at the local $\lambda_I$ makes the intermediate discontinuity a
RD. By comparing (c)
and (d), we note that the potential minimum is exactly the location of
${\bf q}_m$ as expected and it is below the degeneracy point ($b_z |_D
> b_z |_{{\bf q}_m}$). This fact is consistent with the comment in
section V, which predicts that ${\bf q}_m$ will be inside the
degeneracy band. The horizontal lines ${\bf q}_r-{\bf q}_m$ and ${\bf
q}_m-{\bf b}$ measure the propagation speed of the SS and the RD,
which in this case are both zero in the upstream intermediate
frame. They therefore form a compound SS/RD wave.  The downstream of the slow shock (point ${\bf q}_m$) is not able
to connect to the slow rarefaction (SR) wave (point ${\bf a}$) and
thus not able to form a compound SS/SR, since the rarefaction is
faster than the shock itself. This model gives an theoretical
explanation for the possible satellite observations of compound SS/RD
\cite{whang96a, whang97a}, and the ``compound SS/RD/SS waves'' seen in
hybrid simulations
\cite{LCLee00a}.

When $V_S \gtrsim 0$ the potential tilts down in the negative $b_z$
direction (see Fig.~\ref{potential_IS}). In this case, the slow shock
(point ${\bf q}_r$ to ${\bf q}_m$ in Fig.~\ref{potential_IS}(d)) with
shock speed $V_S$ is connected by an intermediate shock (IS) (point
${\bf q}_m$ to ${\bf b}$) with shock speed $V_S$, whose upstream is
super-intermediate ($V_S > \lambda_{I}|_{{\bf q}_m}$) while the
downstream is sub-intermediate ($V_S < \lambda_{I}|_{\bf b}$). This
forms a compound SS/IS wave. Note that the intermediate shock is not
steepened from the intermediate mode (which is consistent with the
discussion in Sec. IV), but is steepened from the slow mode.  The slow
shock is abnormal with both upstream (point ${\bf q}_r$) and
downstream (point ${\bf q}_m$) being super-intermediate. Karimabadi
{\it et al.} \cite{karimabadi95a} call this kind of slow shock an
anomalous slow shock. When $V_S \lesssim 0$ the potential tilts up in
the negative $b_z$ direction and there is no extra transition at the
SS downstream, since $\Psi({\bf b})> \Psi({\bf q}_m)$ in this case and
is therefore not accessible.

These results are independent of the details of $\varepsilon(b_t)$,
but only require the reversal behavior somewhere downstream of the
slow shock. This fact can be inferred from a simple relation:
$V_S-\lambda_I(b_z)=(b_{z0}/b_z)V_S$ for $b_z=b_z|_{{\bf q}_m}$ or
$b_z|_{\bf b}$, regardless of the detail of $\varepsilon(b_t)$ (from
Eq.~(\ref{Fz}) and (\ref{lambda_I})). When $V_S=0$, this relation
ensures that the SS can always connect to a RD since $V_S-\lambda_I=0$
at both points ${\bf q}_m$ and ${\bf b}$. When $V_S \gtrsim 0$, the SS
can always connect to an IS since $V_S-\lambda_I $ is positive
(super-intermediate) at ${\bf q}_m$ and negative (sub-intermediate) at
${\bf b}$.  We therefore conclude that the abnormal transitions of
magnetic field structures seen in the PIC simulations of Paper I are
most likely the transitions from the SS to the RD in a compound SS/RD
wave or the SS to the IS in an SS/IS wave. We can hardly distinguish
between these two compound waves in our PIC simulation, since $V_S$ is
small and the time-dependent dynamics add uncertainties in measuring
the exact value. We focus on further analyzing the compound SS/RD
wave.
 
\subsection{The significance of $\varepsilon=0.25$}  
  
For SS/RD waves ($V_S=0$), the stationary points along $b_{z,s}$ are
the roots of $F_z=0$,
\begin{equation}
\alpha_{\text{eff}}(b_{z,s}) b_{z,s}(b_{z,s}^2-b_{z0}^2)=0
\end{equation}
Here the subscript ``s'' represents ``stationary''. We have three
traditional stationary points, $b_{z,s}=b_{z0}$ (point ${\bf q}_r$),
$-b_{z0}$ and $0$, as well as a new stationary point due to the
temperature anisotropy, $b_{z,m}$ (the transition point ${\bf q}_m$)
determined by $\alpha_{\text{eff}}(b_{z,m})=0$. The fixed-point
analysis of the first three points in isotropic fluid theory can be
found in literature \cite{kennel90a, wu92a, hau89a, hau90a}.

As shown in Fig.~\ref{LO_noSS}, there is no slow mode transition if
\begin{equation}
\varepsilon_0 < \frac{3\gamma-4}{3\gamma-1}=0.25 
\end{equation}
(with $\gamma=5/3$ for monatomic plasma; note that this relation is
independent of $\theta_{BN}$ and $\beta$), which occurs when the
nonlinearity $\alpha$ of Eq.~(\ref{ADNLSB}) changes sign from
$\alpha<0$ to $\alpha>0$. A positive $\alpha$ will result in a
positive $\alpha_{\text{eff}}$ in Eq.~(\ref{alpha_eff}), and therefore
no solution for $b_{z,m}$. Only rotation of the magnetic field is thus
allowed.  If $\varepsilon_0 > 0.25$, we can further show that
$\varepsilon_0 \geq \varepsilon_m(\equiv \varepsilon|_{{\bf q}_m})
\geq 0.25$ is always true for a slow shock transition from
$\varepsilon_0$ to $\varepsilon_m$ in this compound wave by the full
jump conditions of anisotropic-MHD (Appendix IV. B). Therefore, the
nonlinear fluid theory provides a lower bound of $\varepsilon_m \geq
0.25$ at the SS to RD transition inside these compound waves,
regardless of the details of $\varepsilon(b_t)$. In other words, the
downstream magnetic field cannot exhibit switch-off behavior if the
firehose-sense temperature anisotropy is strong enough. This fact
explains the non-switch-off slow shocks often seen in kinetic
simulations \cite{lottermoser98a} and satellite crossings
\cite{seon96a}. Once it transitions to the intermediate mode, a
gyrotropic $\varepsilon(b_t)$ will stay close to $\varepsilon_m$,
since the intermediate-rotation nearly preserves the magnitude of the
$B$ field and therefore $\varepsilon$. Note the assumption of gyrotropic 
$\varepsilon(b_t)$ is expected to be valid only in length scale larger than local
ion inertial length and ion gyro-radius. 

In these demonstrations we use shocks with moderate
parameters, such as $\theta_{BN}=42^\circ$ and $\beta_0=1$. In
general, larger $\theta_{BN}$, $\beta_0$, and smaller $\varepsilon_0$
will make the ratio $|\Omega|/|\alpha|$ larger, and therefore generate
a stronger reversal tendency. An analysis with full anisotropic MHD
(Appendix IV) should be used for strong slow shock transitions, due to
the limits of the ADNLSB, although the underlying physical picture
will be similar.

\section{Toward the critical $\varepsilon=0.25$: time-dependent dynamics}\

The initial conditions that characterize the exhaust of anti-parallel
reconnection (initial $b_y=0$) require ${\bf b}_t=0$ at the symmetry
line at later time. This eventually forces the pseudo-particle to
climb up the potential hill to the local maximum $(b_z=0,b_y=0)$,
which implies an intrinsic time-dependent process at the symmetry line
since Eq.~(\ref{stationary}) does not yield such a
solution. Meanwhile, the fact that ${\bf b}_t$ needs to go to zero at
the symmetry line provides a spatial modulation on the amplitude of
the rotational intermediate mode. Note that the transition point from
SS to RD in compound SS/RD waves could potentially induce modulation
too. As suggested in Paper I, a spatially modulated rotational wave
tends to break into $d_i$-scale dispersive waves, which can make the
rotational component of the transition very turbulent.
 
As pointed out in Sec VI.~B, the nonlinear fluid theory of the
time-independent stationary solutions only provides a lower bound
$\varepsilon_m > 0.25$ for the transition point inside these compound
SS/RD waves.  Counter-streaming ions, by raising $P_\|$, push
$\varepsilon_m$ lower. Once $\varepsilon_m$ is lower than 0.25, the
magnetic field rotates, generates $d_i$-scale waves, and scatters
$P_\|$ into $P_\perp$. This raises $\varepsilon_m$, changing the
functional form of $\delta \varepsilon$ and driving it toward 0, which
self-consistently results in a transition at the potential minimum
where $\alpha_\text{eff} = \alpha=0$, and thus
$\varepsilon=0.25$. This argument explains the $\varepsilon=0.25$
plateau observed in the PIC simulations for different shock angles
(see Fig.~\ref{LO_epsilons}). With $\delta
\varepsilon=0$, this point is exactly the degenerate point of the slow
and intermediate modes.

\section{The supporting evidence from numerical experiments}\

The evidence for a slow mode connecting to rotational waves can be
seen in the PIC simulations that are discussed in detail in Paper
I. In previous kinetic simulations, the downstream rotational waves
were often identified as slow dispersive wavetrains arising from, for
instance, the second term in the RHS of Eq.~(\ref{ADNLSB}). Here we
present further evidence, in addition to the numerical evidence that
$\varepsilon=0.25$, to support the idea that the downstream rotational
mode is tied to the intermediate mode. Fig.~\ref{75d_proves} shows the results from three PIC
simulations that were designed to explore the structure of
reconnection exhausts in the normal direction. The format is the same
as for Fig.~5 of Paper I, with the first column showing $\varepsilon$,
the second column the magnetic field components, and the third column
a hodogram of the fields. The dashed curves in the second column (from
ideal isotropic MHD \cite{lin93a}) indicate that a pair of
switch-off slow shocks or a pair of rotational discontinuities will
propagate out from the center. All three cases show the correlation
between $\varepsilon=0.25$ and the transition from coplanar to
non-coplanr rotation of the downstream magnetic fields. The hodograms
are readily comparable to the state space plots such as
Fig.~\ref{degeneracy} (c). In Fig.~\ref{75d_proves}(a), the downstream
region of a slow shock shows a high wavenumber ($\sim 6 d_i$)
left-handed (LH) polarized rotational wave, which is difficult to
distinguish from the predicted downstream ion inertial scale
dispersive slow mode wavetrain
\cite{coroniti71a}. Fig.~\ref{75d_proves}(b) shows results from a
simulation with a larger initial current sheet width and exhibits a
longer wavelength ($\sim 30 d_i$) LH rotational wave which can be
identified as an intermediate mode 
%(also the only possible LH mode in
%the reversal region is an intermediate mode as predicted from
%anisotropic MHD linear theory. \cite{walthour94a}). 
The intermediate mode breaks into smaller ion inertial scale waves,
which have been identified as dispersive waves in Paper I. By
comparing (a) and (b), we note that the downstream primary rotational
wave tends to maintain its spatial scale as an intermediate mode with
non-steepening and non-spreading properties. Another way to
distinguish the dispersive behavior from the non-dispersive rotation
is by including a weak guide field. In Fig.~\ref{75d_proves}(c), the
front of the rotational downstream wave turns into a well-defined RD
when a weak guide field is included. Its amplitude is about the same
as that of the large amplitude rotational waves in (a). Most
importantly, there is a clear slow shock ahead of the RD. Because the
symmetry of the initial condition is broken by the guide field, the
downstream RD does not need to end at $b_y=0$; instead it ends inside
the potential valley at ${\bf b}_t=(0,b_{z,m})$ (see the hodogram of
(c)), as expected.

\section{Conclusion and discussion}\
  
The existence of compound SS/RD, SS/IS waves arising from
firehose-sense and $|B|$-correlated $\varepsilon$ (temperature
anisotropies) are theoretically demonstrated by analyzing the
anisotropy-caused reversal of a pseudo-potential. The pseudo-potential
is known to characterize the nonlinearity of hyperbolic wave
equations. Extra degeneracy points between slow and intermediate modes
as well as extra non-convexity points in the slow characteristics are
introduced by the temperature anisotropy. The slow shock portion of a
compound SS/IS wave is an anomalous slow shock with both up and
downstream being super-intermediate. The nonlinear fluid theory
provides a lower bound of $\varepsilon=0.25$ for the SS to RD
transition, regardless of the details of the energy closure
$\varepsilon(b_t)$. The wave generation from the rotational
intermediate mode discussed here and in Paper I helps keep
$\varepsilon=0.25$. This explains the critical anisotropy plateau
observed in the oblique slow shock PIC simulations documented in Paper
I. This study also suggests that it is a pair of compound SS/RD waves
that bound the antiparallel reconnection outflow, instead of a pair of
switch-off slow shocks as in Petschek's reconnection model. This fact
explains the in-situ observations of non-switch-off slow shocks in
magnetotail \cite{seon96a}. It also provides a theoretical explanation
of the observations of ``Double Discontinuity'' by Whang {\it et al.}
\cite{whang96a,whang97a} with GEOTAIL data, and also the step-like
slow shocks seen in Lottermoser {\it et al.}'s large-scale hybrid
reconnection simulation
\cite{lottermoser98a}. In previous hybrid and PIC simulations, the downstream sharp
rotational waves were often identified as slow dispersive waves of a
switch-off slow shock. Instead, we propose that they are the
intermediate portion of the compound SS/RD wave. The slow shock
portion becomes less steep due to the time-of-flight effect of
backstreaming ions.

%As pointed out in Sec.~VIII, due to the symmetry
%of the antiparallel asymptotic magnetic fields, the reconnection
%downstream has an intrinsically time-dependent nature sourced from the
%symmetry line.

The singularity of $\varepsilon=0.25$ was also noticed by P.~D.~Hudson
\cite{hudson71a} in his study on the anisotropic rotational
discontinuity (A-RD). Unlike the RD in a compound SS/RD wave, an A-RD
changes both $\varepsilon$ and thermal states. Through the constraint
of the positivity of $P_\|$, $P_\perp$ and $B^2$, he derived all of
the possible jumps (independent of the energy closure used) of the
temperature anisotropy across an A-RD, as shown in
Fig.~\ref{hudson}(a). (Note that in Fig.~\ref{hudson} $\Delta
\varepsilon=\varepsilon_\text{up}-\varepsilon_\text{down} \rightarrow
0$ when $\varepsilon \rightarrow 0.25$.) The ADNLSB inherits most of
the hyperbolic properties (such as the extra nonconvexity and
degeneracy points) of anisotropic MHD and also this singular
behavior. We can tell this by searching for stationary solutions where
the pseudo-force $F_z=0$ (the general form of Eq.~(\ref{Fz})). Again,
\begin{equation}
V_S(b_z-b_{z0})-\alpha b_z(b_z^2-b_{z0}^2)-\Omega b_z(\varepsilon-\varepsilon_0)=0
\end{equation}
A stationary A-RD exists at $\varepsilon_0= 0.25$ ({\it i.e.,}
$V_S=0$, $\alpha=0$ and $b_z \neq 0$), only if we require
$\varepsilon-\varepsilon_0=0$. An arbitrary magnetic field magnitude
and rotation are then allowed, as shown by Hudson. After further
constraining the possible jumps by requiring that entropy increase,
the solution above the diagonal line in Fig.~\ref{hudson}b is
eliminated when $\varepsilon_\text{up} > 0.25$ while the solution
below the diagonal line is eliminated when $\varepsilon_\text{up} <
0.25$. He also noticed that the jump behavior of an A-RD for
$\varepsilon_\text{up} > 0.25$ is slow-mode-like ({\it i.e.}  $\delta
n$ and $\delta B$ are anti-correlated), while it is fast-mode-like
({\it i.e.} $\delta n$ and $\delta B$ are correlated) for
$\varepsilon_\text{up} < 0.25$. This directly relates to the fact that
the jump of an A-RD equals the jump of the compound SS/RD wave (see
Appendix IV.~C), and a slow mode turns fast-mode-like when $\alpha >
0$ \cite{kennel90a}.

\section{Appendices}

\textbf{I - From anisotropic MHD to the Anisotropic DNLSB equation:}

From moment integrations of the Vlasov equation that neglect the
off-diagonal components of the pressure tensor (the empirical validity
of this approximation for our system is shown in Fig. 3(c) of Paper
I), we can write down the anisotropic MHD (AMHD) equations
\cite{chao70a,hudson70a}. The energy closure is undetermined.

In Lagrangian form,
\begin{equation}
\frac{d}{dt} \rho +\rho \partial_x V_x=0
\end{equation} 
\begin{equation}
\rho \frac{d}{dt} V_x +\partial_x P+\partial_x
\left[\frac{1}{3}\left(\varepsilon+\frac{1}{2}\right)\frac{B_t^2}{\mu_0}\right]
-\frac{2}{3}\frac{B_x^2}{\mu_0}\partial_x
\varepsilon =0
\end{equation} 
\begin{equation}
\rho \frac{d}{dt} {\bf V}_t -\frac{B_x}{\mu_0}\partial_x(\varepsilon {\bf B}_t)=0
\end{equation} 
\begin{equation}
\frac{d}{dt} {\bf B}_t -B_x\partial_x{\bf V}_t+{\bf B}_t\partial_x
V_x=\partial_x (\eta_r \partial_x {\bf B}_t)
-\partial_x\left(\frac{B_x}{\mu_0 ne}\hat{\bf e}_x\times \partial_x {\bf
B}_t\right)
\label{original}
\end{equation} 
\begin{equation}
\frac{d}{dt} {P} -\frac{\gamma
P}{\rho}\frac{d}{dt}\rho+(\gamma-1)\left[\frac{1}{3}(\varepsilon+2)\frac{B^2}{\mu_0}-\varepsilon
\frac{B^2_x}{\mu_0}-\frac{B^2_t}{\mu_0}\right]\partial_x V_x
+(\gamma-1)(1-\varepsilon)\frac{B_x{\bf B}_t}{\mu_0}\cdot \partial_x
{\bf V}_t +(\gamma-1)\partial_x Q_x=0
\end{equation} 
where
\begin{align}
\varepsilon \equiv 1-\frac{P_\|-P_\bot}{B^2/\mu_0},  \qquad  P \equiv \frac{P_\|+2P_\bot}{3}.
\end{align} 
$\gamma =5/3$ or $7/5$ for monatomic or diatomic plasma, respectively
\cite{hau02a}. $\varepsilon$, $P_\|$, $P_\bot$, $\rho$, $V_x$, ${\bf
V}_t$, $B_x$, ${\bf B}_t$, $Q_x$ and $\eta_r $ are the temperature
anisotropy factor, pressure parallel to the local magnetic field,
pressure perpendicular to the local magnetic field, mass density,
velocity of the bulk flow in the normal direction ($\hat{{\bf e}}_x$),
velocity of the bulk flow in the tangential direction ($y$-$z$ plane), 
normal component of the magnetic field, tangential components
of the magnetic field, the heat flux in the $x$-direction and the
magnetic resistivity (assumed constant). The first and the second term
on the RHS of Eq.~(\ref{original}) are from the magnetic dissipation
and the Hall term respectively.
   
Then we follow the procedure of Kennel {\it et al.}
\cite{kennel90a}, using Lagrangian mass spatial coordinates,

\begin{align}
\frac{d}{dt} \rightarrow \partial_t,  \qquad  x' \equiv \int{\frac{\rho}{\rho_0}dx}
\end{align} 

Jumping to the upstream (subscripted by ``0'') intermediate frame in
order to separate the slow and fast variations, we take
\begin{align}
\tau \equiv \sqrt{\varepsilon_0}C_{An}t,  \qquad  \eta \equiv x'-\sqrt{\varepsilon_0}C_{An}t
\end{align} 
where $C_{An}^2 \equiv B_x^2/(\mu_0 \rho_0)$, and with the approximations
\begin{align}
\partial_\tau \ll \partial_\eta; && \delta b_t^2 \ll b_{t0}^2; && \delta \varepsilon \ll \varepsilon_0; && \delta \rho \ll \rho_0; && \delta P \ll P_0,
\end{align}  
(where $\delta$ means variation), we can collapse the original seven
equations into two coupled equations
\begin{equation}
\partial_\tau {\bf b}_t + \partial_\eta[\alpha{\bf b}_t
(b_t^2-b_{t0}^2)+\Omega{\bf b}_t(\varepsilon-\varepsilon_0)+\Lambda
{\bf b}_t(Q_x-Q_{x0})] =\partial_\eta(R\partial_\eta {\bf
b}_t)-\frac{1}{2\sqrt{\varepsilon_0}}d_i\partial_\eta^2(\hat{\bf e}_x
\times {\bf b}_t)
\end{equation}
with
\begin{align}
\alpha\equiv\frac{[(3\gamma-1)\varepsilon_0-(3\gamma-4)]C_{An}^2}{12[(\varepsilon_0-A)C_{An}^2-C_S^2]};
&&
\Omega\equiv\frac{(b_{t0}^2-2)C_{An}^2}{6[(\varepsilon_0-A)C_{An}^2-C_S^2]}+\frac{1}{2\varepsilon_0};
&& \Lambda\equiv \frac{(\gamma-1)
C_{An}^2}{2\sqrt{\varepsilon_0}[(\varepsilon_0-A)C_{An}^2-C_S^2]}
\end{align}
and
\begin{equation}
A\equiv\frac{2}{3}(\gamma-1)(1-\varepsilon_0)(1+b_{t0}^2)
\end{equation}
where $C_S^2 \equiv \gamma P_0/\rho_0$, $R\equiv
\eta_r/(2\sqrt{\varepsilon_0}C_{An})$, and $d_i\equiv \sqrt{m_i/(\mu_0 n
e^2)} $.

Since the heat flux $Q_x$ is approximately proportional to $\nabla
|B|$ (as pointed out in Fig.~3(a) of Paper I), it should enter the
source term on the RHS as $-\Lambda \partial_\eta^2 |b_t|$. In plasmas
with $\beta_n>1$, $\Lambda$ is negative and hence the heat flux helps
shocks dissipate energy. We implicitly incorporate it into the
resistivity $R$ (as, for instance, is done for the shear and
longitudinal viscosities discussed in \cite{kennel90a}). We then
arrive at the anisotropic DNLSB equation, Eq.~(\ref{ADNLSB}).  This
equation can also be derived from regular reductive perturbation
methods with a proper ordering scheme.  For instance, a DNLS equation
with the CGL condition and more corrections, including finite ion
Larmor radius effects and electron pressure, was derived using regular
reductive perturbation methods \cite{khanna82a}.

\textbf{II - Non-steepening and non-spreading of the intermediate mode:} 

Beginning with the anisotropic MHD equations in Lagrangian form in
Appendix I, we neglect the dissipation and the Hall term on the RHS of
Eq.~(\ref{original}).  Then the simple wave solution can be obtained
by substituting $d/dt \rightarrow -C \delta$, $\partial_x \rightarrow
\hat{{\bf e}}_x \delta$, where $C$ is the wave speed and $\delta$
means variation. \cite{jeffrey64a}
\begin{equation}
-C\delta \rho+\rho \delta V_x=0
\label{continue}
\end{equation} 
\begin{equation}
-C\rho \delta V_x + \delta P_\perp- \frac{B_x^2}{\mu_0}\delta
\varepsilon+ \frac{B_z}{\mu_0} \delta B_z=0
\label{normal_m}
\end{equation} 
\begin{equation}
-C\rho \delta V_z - \varepsilon \frac{B_x}{\mu_0}\delta B_z-
\frac{B_xB_z}{\mu_0}\delta \varepsilon=0
\label{transverse_m_in}
\end{equation} 
\begin{equation}
-C\rho \delta V_y - \varepsilon \frac{B_x}{\mu_0}\delta B_y=0
\label{transverse_m_out}
\end{equation} 
\begin{equation}
-C \delta B_z+ B_z \delta V_x- B_x \delta V_z =0
\label{ohms_in}
\end{equation} 
\begin{equation}
-C \delta B_y-B_x \delta V_y =0
\label{ohms_out}
\end{equation} 

Eq.~(\ref{transverse_m_out}) and Eq.~(\ref{ohms_out}) give us the
intermediate speed $C_I=\sqrt{\varepsilon}B_x/\sqrt{\mu_0
\rho}$. Combined with Eq.~(\ref{continue}), the steepening tendency of
an intermediate mode can then be expressed as,
 
\begin{equation}
\delta (C_I+ V_x)=\frac{C_I}{2}\left(\frac{\delta \rho}{\rho}+\frac{
\delta \varepsilon}{\varepsilon}\right)
\end{equation} 
Using Eqs.~(\ref{continue}), (\ref{normal_m}), (\ref{transverse_m_in})
and (\ref{ohms_in}), we get $\delta(C_I+ V_x)=0$. Therefore, the
intermediate mode in anisotropic MHD does not steepen or spread, no
matter what energy closure is used. It is linearly degenerate, as is
its counterpart in isotropic MHD.

\textbf{III -The integral curves and Hugoniot Locus:}

To find the integral curves, we follow the eigenvector of the slow
mode to form a curve,
\begin{align}
\frac{db_z}{d \zeta}=b_z,  \qquad  \frac{db_y}{d\zeta}=b_y,
\end{align} 
where $\zeta$ is a dummy variable. The integral curve is $b_y=(b_{y0}/b_{z0})b_z$.

For the intermediate mode,
\begin{align}
\frac{db_z}{d \zeta}=b_y,  \qquad  \frac{db_y}{d\zeta}=-b_z.
\end{align} 
Therefore, the integral curve is $b_t^2=b_{t0}^2$.

As to the Hugoniot locus, we need to compute the shock speed
$S=(f_i({\bf q})-f_i({\bf q}_0))/(q_i-q_{0i})$ where $f_i$ is the flux
of Eq.~(\ref{flux}) and $i=1$ or $2$.
\begin{equation}
\begin{split}
S(b_z-b_{z0})=\alpha b_z(b_t^2-b_{t0}^2)+\Omega b_z\delta\varepsilon\\
S(b_y-b_{y0})=\alpha b_y(b_t^2-b_{t0}^2)+\Omega b_y\delta\varepsilon
\end{split}
\end{equation} 
These can be combined to give,
\begin{equation}
(b_{z0}b_y-b_{y0}b_z)[\alpha (b_t^2-b_{t0}^2)+\Omega \delta\varepsilon]=0.
\end{equation} 
The first root is the Hugoniot locus of the slow mode:
$b_y=(b_{y0}/b_{z0})b_z$.  For $\varepsilon(b_t)$, so that
$\delta\varepsilon \simeq [\partial(\delta
\varepsilon)/\partial(b_t^2)](b_t^2-b_{t0}^2)$, the second root gives
us the Hugoniot locus of the intermediate mode:
$b_t^2=b_{t0}^2$. Although these results are the same as derived from
the integral curves, this is not generally the case.

\textbf{IV - The pseudo-potential of Anisotropic MHD (AMHD):} 

In the de Hoffmann-Teller frame, the jump conditions can be written as
(following Hau and Sonnerup's procedure \cite{hau89a, hau90a}),
\begin{equation}
\left[ \rho V_x \right]^0=0
\label{continue_J}
\end{equation} 
\begin{equation}
\left[\rho
V_x^2+P+\frac{1}{3}\left(\varepsilon+\frac{1}{2}\right)\frac{B^2}{\mu_0}-\varepsilon
\frac{B_x^2}{\mu_0} \right]^0=0
\label{normal_m_J}
\end{equation} 
\begin{equation}
\left[\rho V_x {\bf V}_t -\varepsilon \frac{B_x{\bf B}_t}{\mu_0} \right]^0=0
\label{transverse_m_J}
\end{equation} 
\begin{equation}
\left[\left(\frac{1}{2} \rho V^2 + \frac{\gamma}{\gamma-1}P
+\frac{1}{3}(\varepsilon-1)\frac{B^2}{\mu_0}\right)V_x-(\varepsilon-1)
\frac{B_x{\bf B}_t}{\mu_0} \cdot {\bf V}_t-(\varepsilon-1)
\frac{B_x^2}{\mu_0}V_x \right]^0=0
\label{energy_J}
\end{equation} 
where we define a jump relation $[Q]^0\equiv Q_0-Q$, with $Q_0$ the
upstream value and $Q$ the value inside the transition region. From
Eq.~(\ref{continue_J})-(\ref{energy_J}), we can derive
\begin{equation}
A_x^2 \equiv \frac{V_x^2}{B_x^2/(\mu_0 \rho)}=\frac{-b \pm \sqrt{b^2-4ac} }{2a}
\label{Ax2}
\end{equation} 
where
\begin{equation}
a=1-\frac{\gamma-1}{2\gamma},
\end{equation} 
\begin{equation}
b=-A_{x0}^2-\left[\frac{\beta_0}{2}+\frac{1}{3}\left(\varepsilon_0+
\frac{1}{2}\right)\right]\mbox{sec}^2\theta_0 +
\varepsilon_0+\frac{1}{3}\left(\varepsilon+\frac{1}{2}\right)\frac{B^2}{B_x^2}
-\varepsilon
+\frac{2\gamma-2}{3\gamma}(\varepsilon-1)-\frac{\gamma-1}{3\gamma}(\varepsilon-1)\frac{B_t^2}{B_x^2},
\end{equation} 
\begin{equation}
\begin{split}
c=\frac{\gamma-1}{2\gamma}\mbox{sec}^2\theta_0A_{x0}^4+\left[\frac{\beta_0}{2}-
\frac{2\gamma-2}{3\gamma}(\varepsilon_0-1)\right]\mbox{sec}^2\theta_0
A_{x0}^2-\frac{\gamma-1}{2\gamma}\left[(A_{x0}^2-\varepsilon_0)\mbox{tan}\theta_0+\varepsilon
\frac{B_t}{B_x}\right]^2\\ +
\frac{\gamma-1}{\gamma}(\varepsilon-1)\frac{B_t}{B_x}\left[(A_{x0}^2-
\varepsilon_0)\mbox{tan}\theta_0+\varepsilon
\frac{B_t}{B_x}\right]
\end{split}
\end{equation} 
with $\cos\theta_0 \equiv B_x/B_0$.

The generalized Ohm's law is
\begin{equation}
{\bf E} + {\bf V} \times {\bf B} = \frac{\eta_r}{\mu_0} {\bf
J}+\frac{1}{\mu_0 n e}({\bf J} \times {\bf B}),
\label{ohms}
\end{equation} 
where the first and the second term on the RHS are the magnetic
dissipation and the Hall term respectively. With the final jump
condition $\left[ E_t\right]^0=0$, we obtain
\begin{equation}
\frac{A_{x0}}{h_0}d_{i0}(1+h^2)\frac{dB_y}{dx}=(A_x^2-\varepsilon)(B_y-hB_z)+
(A_{x0}^2-\varepsilon_0)(hB_{z0}-B_{y0})\equiv F_{y,\text{AMHD}}\equiv
\frac{\partial \Psi_{\text{AMHD}}}{\partial B_y}
\label{F_yAMHD}
\end{equation} 
\begin{equation}
\frac{A_{x0}}{h_0}d_{i0}(1+h^2)\frac{dB_z}{dx}=(A_x^2-\varepsilon)(B_z+hB_y)-
(A_{x0}^2-\varepsilon_0)(hB_{y0}+B_{z0})\equiv F_{z,\text{AMHD}}\equiv
\frac{\partial \Psi_{\text{AMHD}}}{\partial B_z}
\label{F_zAMHD}
\end{equation} 
where $h\equiv B_x/ne\eta_r$ measures the ratio of the dispersion to
the resistivity. The pseudo-potential $\Psi_{\text{AMHD}}$ is uniquely
defined since $\partial F_{z,\text{AMHD}}/\partial B_y =
\partial F_{y,\text{AMHD}}/\partial B_z$.

\begin{figure}
\includegraphics[width=18cm]{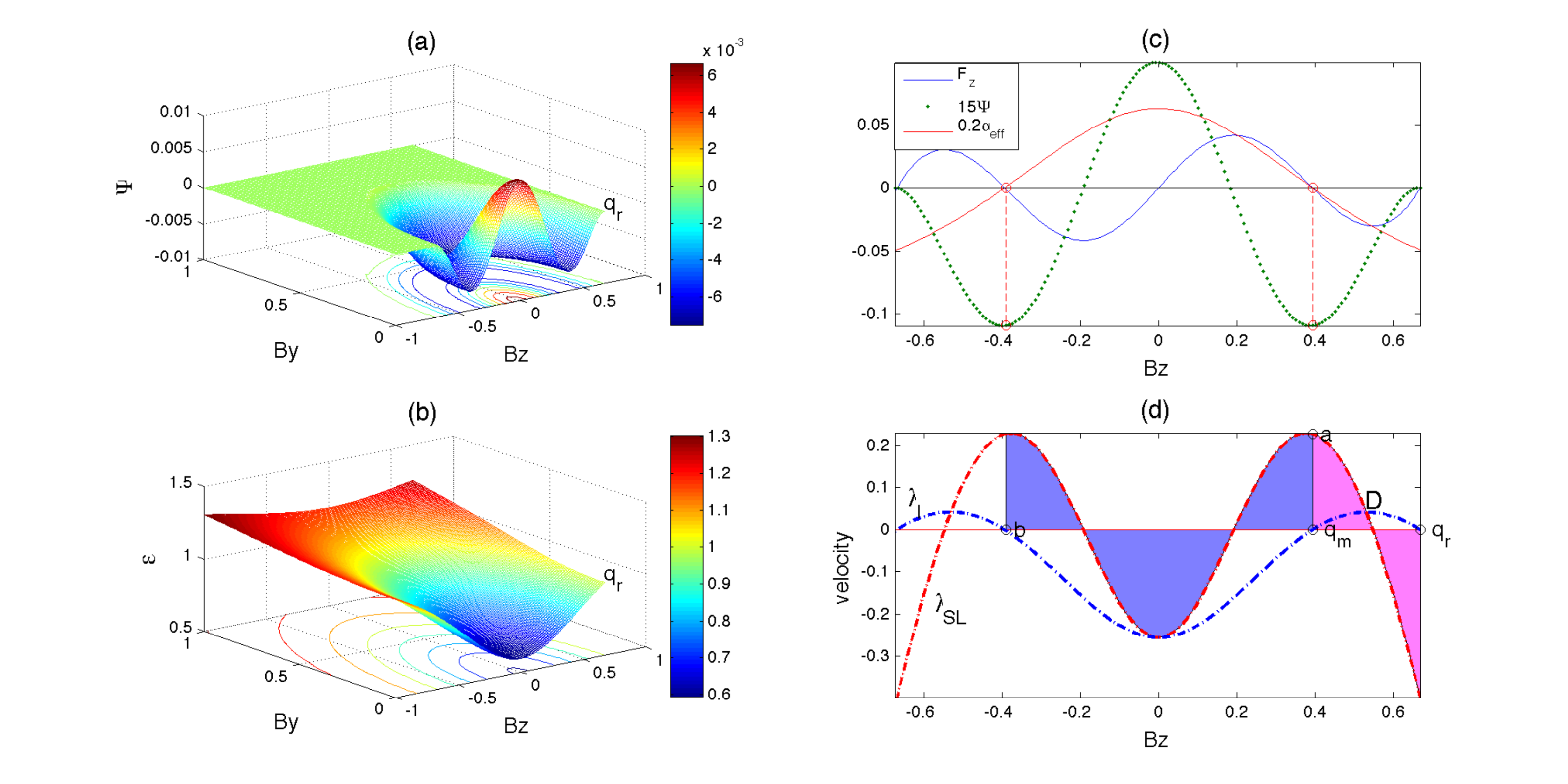} 
\caption{Panel (a): A pseudo-potential $\Psi$ with $V_S=0$. Upstream
(point ${\bf q}_r$), $\theta_0=42^\circ, \beta_0=1,
\varepsilon_0=1,c_2=0.5$ and we choose $\Psi({\bf q}_r)=0$. Since the
transition occurs within the radius $b_t=b_{z0}$, we set $\Psi=0$ for
$b_t > b_{z0}$ for a better visualization. The potential for negative
$B_y$ is mirror symmetric to the part shown here; Panel (b):
$\varepsilon(b_t)$; Panel (c): Cuts of $\Psi$, $F_z$, and
$\alpha_\text{{eff}}$ along the $B_z$ axis with $B_y=0$; Panel (d):
$\lambda_I$ and $\lambda_{SL}$ along the $B_z$ axis with $B_y=0$. The
vertical axis measures speed (normalized to $C_{An}$). ``D'' stands
for degeneracy. The red area above $V_S$ (zero here) equals the red
area below $V_S$, and the same rule applies to the blue area.}
\label{potential} \end{figure}

\begin{figure}
\includegraphics[width=18cm]{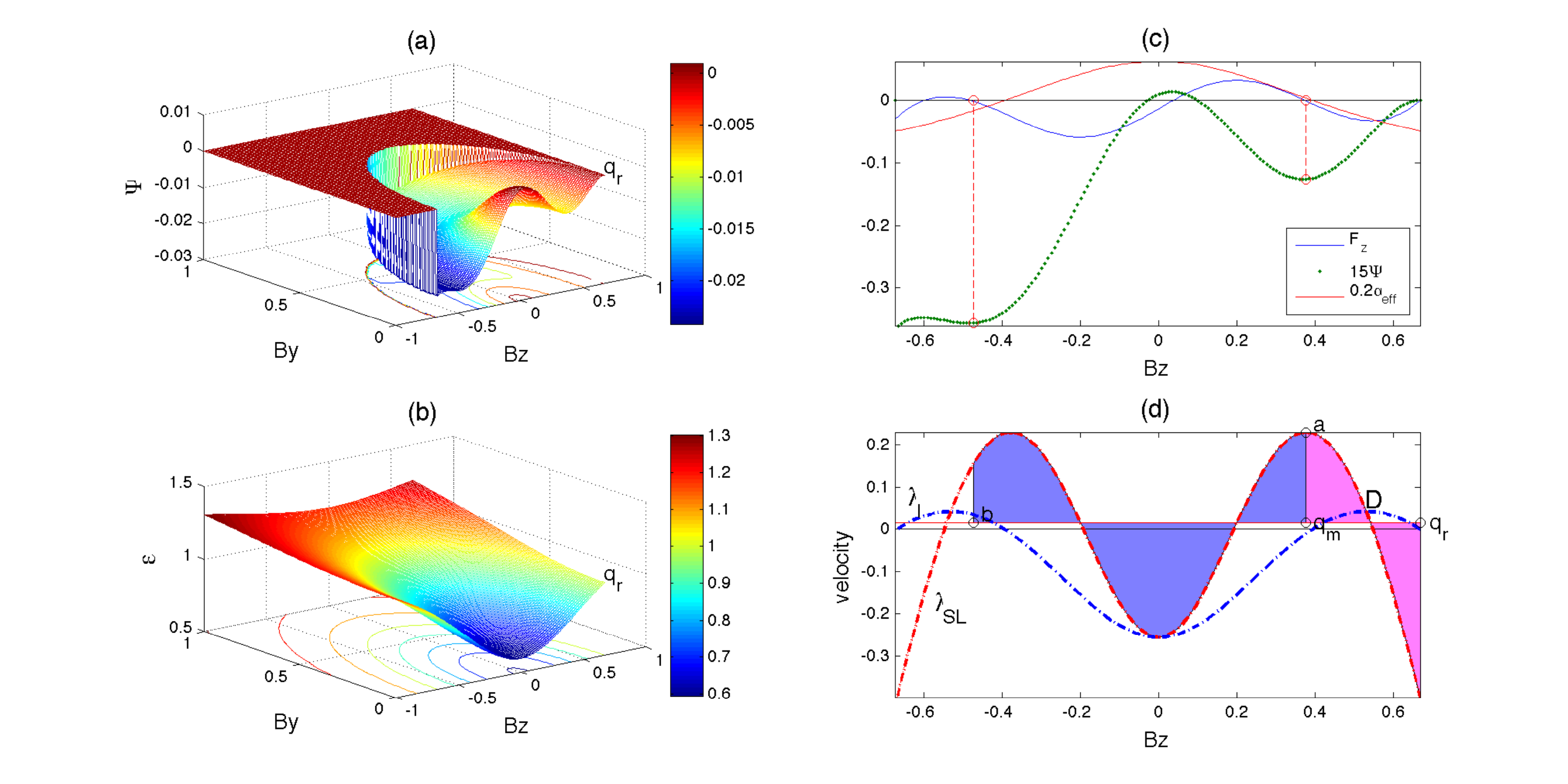} 
\caption{Same format as Fig.~\ref{potential} but with $V_S =0.015 \gtrsim 0$.} 
\label{potential_IS}
\end{figure}

\begin{figure} \includegraphics[width=8cm]{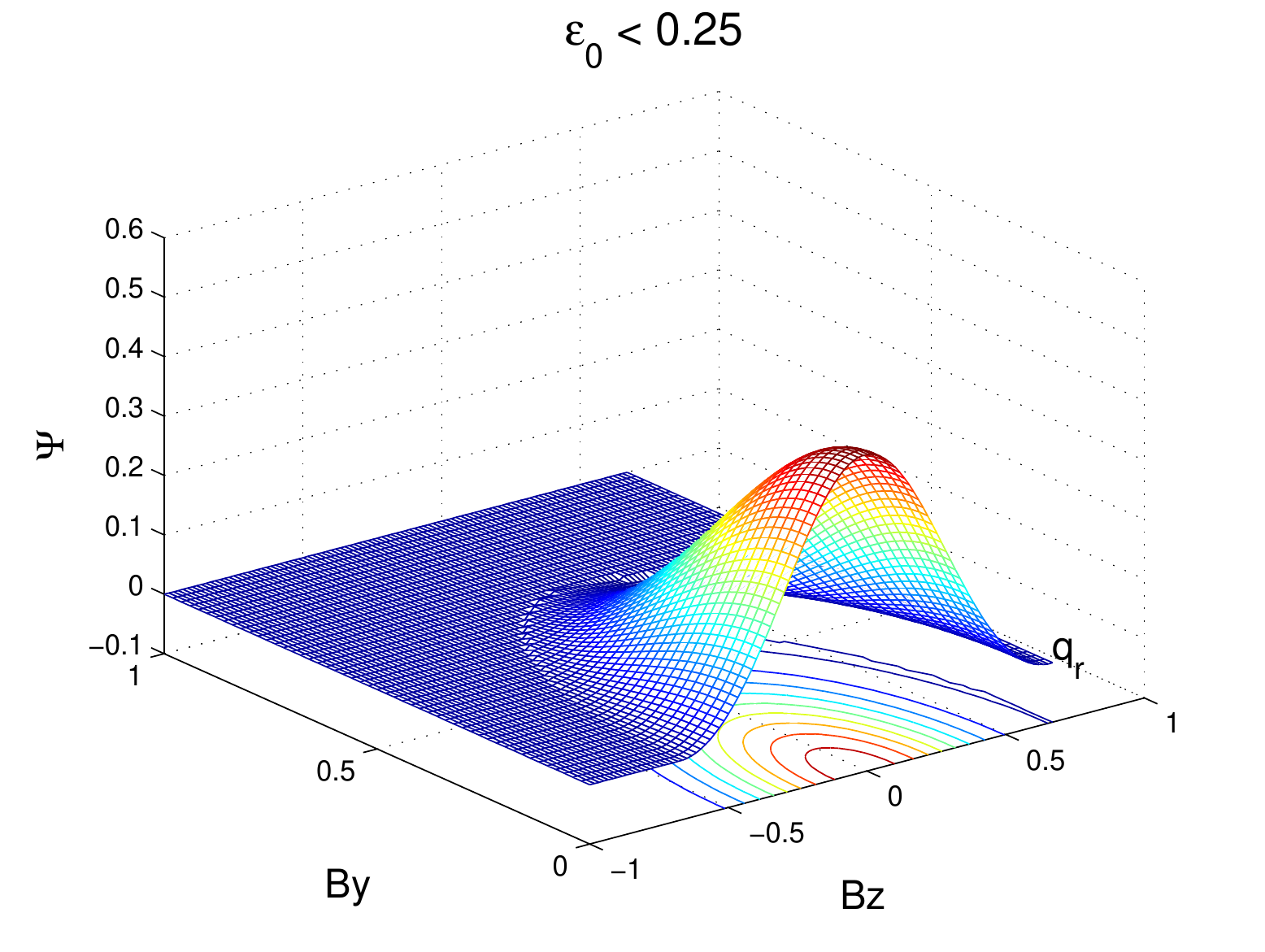} 
\caption{The pseudo-potential with $\varepsilon_0 =0.24 < 0.25$, while
other parameters are the same as Fig.~\ref{potential}.}
\label{LO_noSS}
\end{figure}

\begin{figure}
\includegraphics[width=14cm]{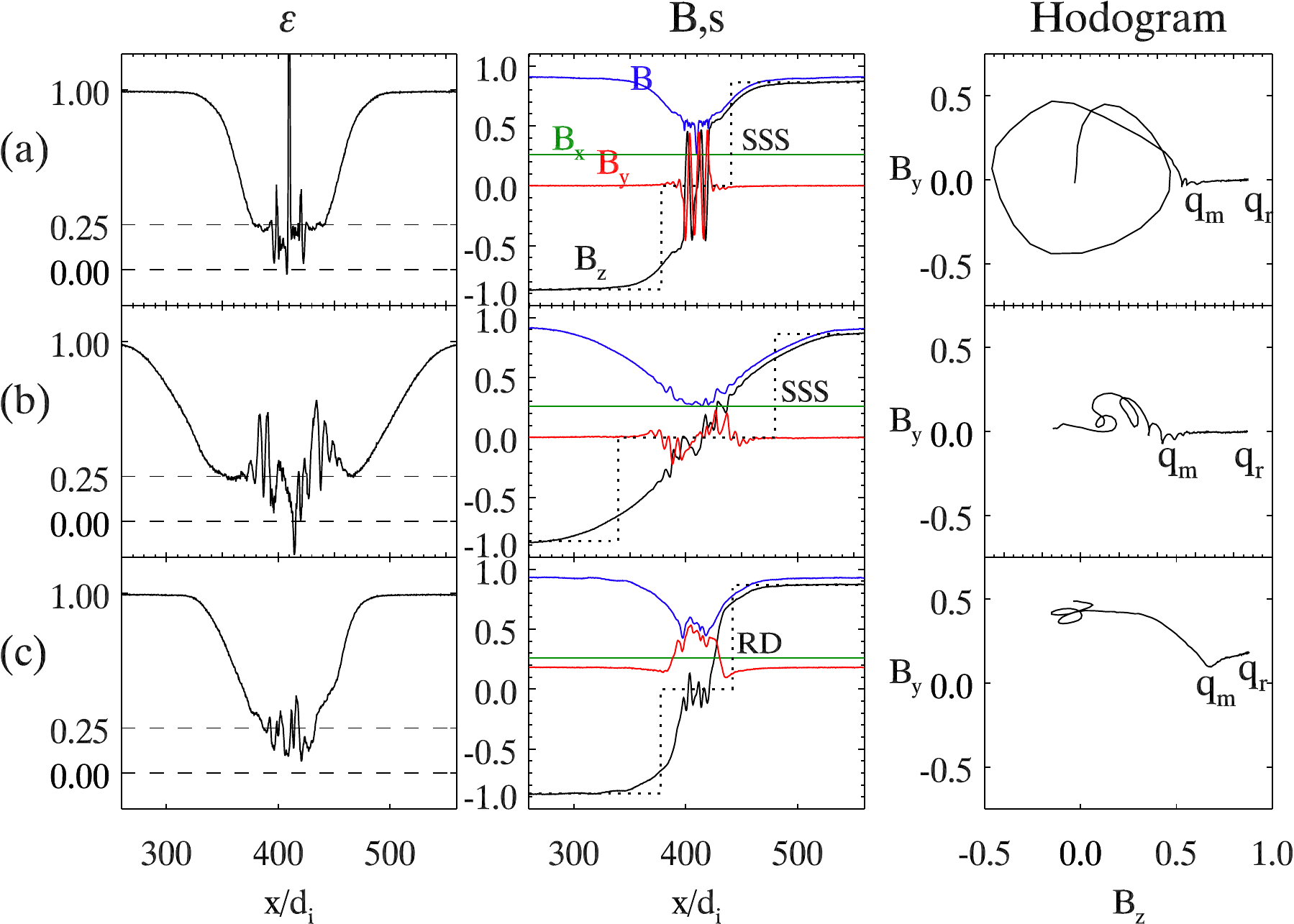} 
\caption{ Results of PIC simulations (runs ${\bf f}$, ${\bf g}$ and
${\bf h}$ of Paper I \cite{yhliu11a}). Row (a): A case with
$\theta_{BN}=75^\circ, \beta_0=0.4$ and initial width $w_i=1d_i$ at
time $200/\Omega_{ci}$. $\varepsilon$ is shown on the left, different
magnetic components in the middle, $B_z$-$B_y$ hodogram on the right;
Row (b): A similar case with a wider initial width $w_i=10d_i$ at time
$450/\Omega_{ci}$; Row (c): A similar case to (a), but with a weak
guide field $B_{y0}= 0.2 B_0$ at time $200/\Omega_{ci}$.  The dotted
curves in the center column are the predicted $B_z $ magnitudes and
positions of switch-off slow shocks (SSS) or rotational
discontinuities (RD) from isotropic MHD theory \cite{lin93a}.}
\label{75d_proves} \end{figure}

\begin{figure}
\includegraphics[width=13cm]{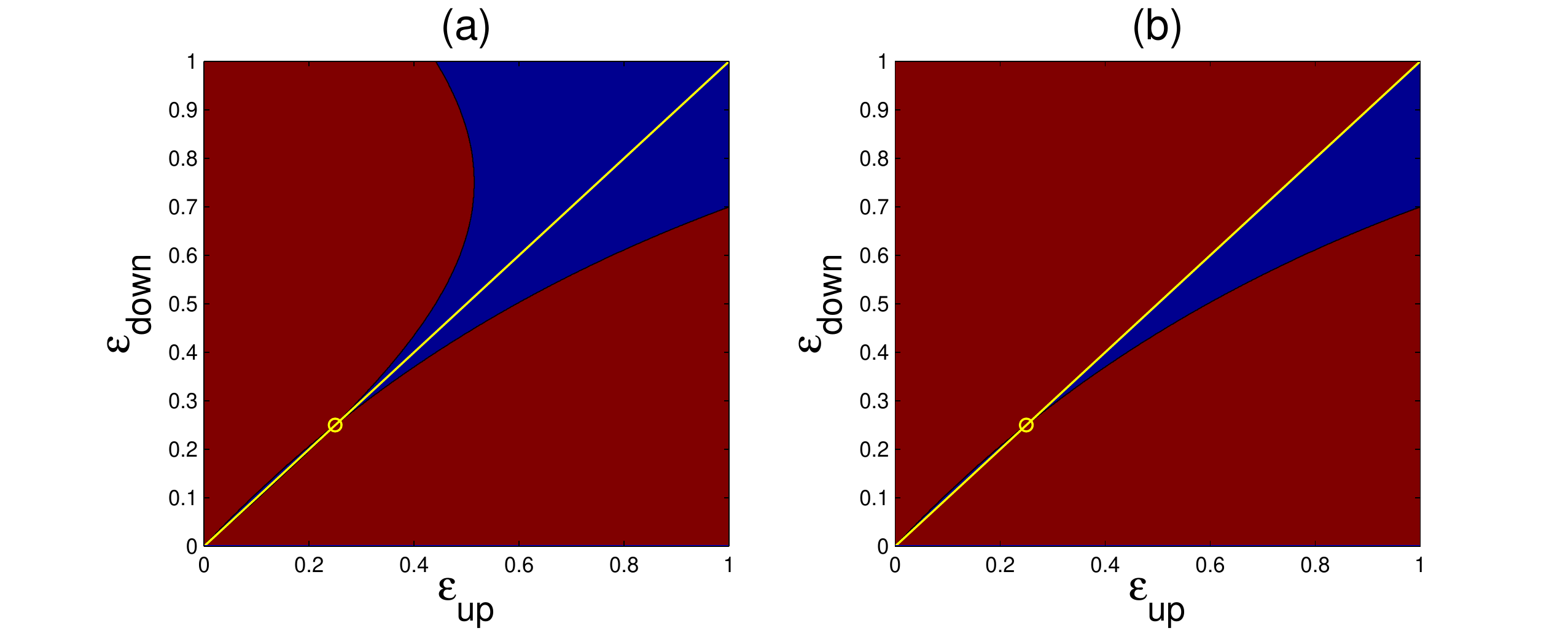} 
\caption{In panel (a), the possible jumps of $\varepsilon$ of an
anisotropic-RD, are constrained by requiring positive $P_{\perp d}$,
$P_{\| d}$ and $B^2_d$. The red region is forbidden. In panel (b), the
plot is further constrained by the requirement that entropy increases
(with an entropy from the H-theorem defined as $\mbox{ln}(
P_\|^{1/3}P_\perp^{2/3}/\rho^{5/3})$ for a bi-Maxwellian
distribution). The constraint of increasing entropy has eliminated the
region above the diagonal line when $\varepsilon_{up}>0.25$, and the
region below the diagonal line when $\varepsilon_{up}< 0.25$. Here
$\beta_0=1$; a higher $\beta_0$ would collapse the valid region into a
narrower region along the diagonal line.}  \label{hudson} \end{figure}

\begin{figure}
\includegraphics[width=16cm]{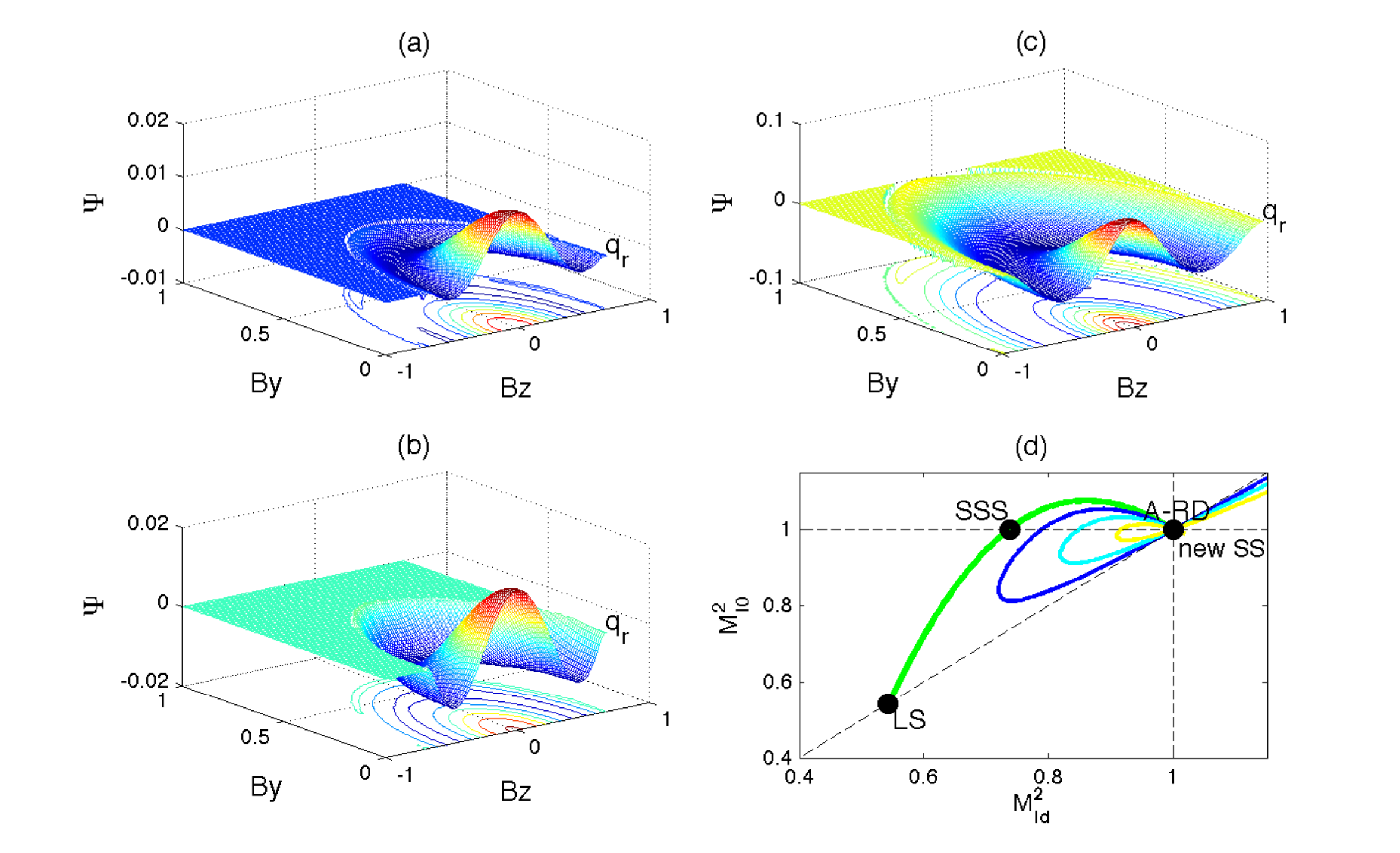} 
\caption{Results with fully anisotropic MHD. Panel (a) is the
pseudo-potential $\Psi$ with our closure,
Eq.~(\ref{closure}). Upstream (point ${\bf q}_r$), $V_S = 0$,
$\theta_0=42^\circ$, $\beta_0=1$, $\varepsilon_0=1$ and $c_2=0.5$;
Panel (b) is $\Psi$ with the CGL closure. Upstream (point ${\bf
q}_r$), $V_S = 0$, $\theta_0=42^\circ$, $\beta_0=1$ and
$\varepsilon_0=1.5$; Panel (c) is $\Psi$ for $V_S = 0$,
$\theta_0=75^\circ$, $\beta_0=0.4$, $\varepsilon_0=1$, and $c_2=0.2$
with our closure; Panel (d) is the shock curve with upstream
parameters $\theta_0=42^\circ$, $\beta_0=1$ and $\varepsilon_0=1$. In
the green curve ($\varepsilon_d=\varepsilon_0$ case), the portion from
A-RD (anisotropic-RD) to SSS is the IS branch, from SSS to LS (linear
slow mode) is the SS branch. Different curves represent cases with
different $\varepsilon_d$ of values 1, 0.95, 0.9, 0.85 (from outer
curve to inner curve). Other than the A-RD, a new SS exists at (1,1)
when $\varepsilon_d < \varepsilon_0$. Both the IS and SS branches
shrink toward the point (1,1) as $\varepsilon_d$ decreases.  }
\label{potential_MHD} \end{figure}

${\bf A}$: Fig.~\ref{potential_MHD}(a) shows the pseudo-potential of
AMHD for the same parameters as Fig.~\ref{potential}(a) (which was
calculated based on the reduced ADNLSB formulation). If $V_S=0$ then
$A_{x0}^2=\varepsilon_0$, and thus the potential minimum (where
$F_{z,\text{AMHD}}=0$) occurs at $A_x^2=\varepsilon$. This implies
that in the shock frame (also the upstream intermediate frame),
$\lambda_{I,\text{AMHD}}=
-V_x+C_I=(-A_x+\sqrt{\varepsilon})B_x/\sqrt{\mu_0 \rho}=0$ at the
potential minimum. This is essentially the same point ${\bf q}_m$
(where $\lambda_I=0$) of Fig.~\ref{potential}(d) with
ADNLSB. Fig.~\ref{potential_MHD}(b) shows a similar reversal with the
CGL closure. We note that the CGL closure exhibits an even stronger
tendency to reverse the pseudo-potential.  In
Fig.~\ref{potential_MHD}(c), the pseudo-potential for
$\theta_{BN}=75^\circ$, $\beta_0=0.4$ and $c_2=0.2$ is shown, these
parameters are more similar to those seen in our PIC simulation.

${\bf B}$: Now we consider the jump conditions to an asymptotic
downstream by neglecting the LHS and terms with $h$ of
Eqs.~(\ref{F_yAMHD}) and (\ref{F_zAMHD}).  The relation
$B_{t,d}/B_x=\mbox{tan}\theta_0(A_{x0}^2-\varepsilon_0)/(A_{x,d}^2-\varepsilon_d)$
can be derived where we label quantities $Q\rightarrow Q_d$ (``d'' for
downstream). We can eventually invert $A_{x0}^2$ as a function of
$A_{x,d}^2$ from Eq.~(\ref{Ax2}). The result is plotted in
Fig.~\ref{potential_MHD}(d) which shows possible shock solutions as
functions of the downstream intermediate Mach number $M_{I,d}^2 \equiv
V_{x,d}^2/C_{I,d}^2=A_{x,d}^2/\varepsilon_d$ \cite{karimabadi95a}. In
the green curve ($\varepsilon_d=\varepsilon_0$ case), the portion from
A-RD (anisotropic-RD) to SSS is the IS branch, from SSS to LS (linear
slow mode) is the SS branch. When $\varepsilon_d < \varepsilon_0$, a
new slow-shock transition from $A_{x0}^2=\varepsilon_0>0.25$ to
$A_{x,d}^2=\varepsilon_d$ is noted at the point (1,1). This new SS
constitutes the slow shock portion of a compound SS/RD wave. For a
given $\varepsilon_0$, the smallest possible
$\varepsilon_{d,\text{min}}$ shrinks the SS and IS branches to the
point (1,1). It can be shown that $\varepsilon_{d,\text{min}} > 0.25$
is always true for $\varepsilon_0>0.25$. Therefore the existence of
the new slow shock requires $\varepsilon_d > 0.25$. In other words,
$\varepsilon_0> \varepsilon_m(=\varepsilon_d)> 0.25$ is always true
for a SS/RD compound wave in full anisotropic MHD.

${\bf C}$: From Fig.~\ref{potential_MHD}(d) and further
investigations, it can be shown that the anisotropic-RD(A-RD) at (1,1)
has the same jump as that of the new SS at (1,1) plus a RD that does
not change $\varepsilon$ and thermal states. Therefore an A-RD and the
corresponding compound SS/RD wave have the same jump relations.

\textbf{V - The Equal-Area Rule and Intermediate Shocks:}

The equal-area rule applies to conserved quantities in hyperbolic
equations, which in our case is $b_z$.  From Eq.~(\ref{flux}) and the
general form of Eq.~(\ref{Fz}), we find a simple relation between the
pseudo-force and the flux function, \begin{equation}
F_z|_{b_y=0}=-\alpha b_z (b_z^2-b_{z0}^2)-\Omega b_z
\delta\varepsilon(b_z,b_y) + V_S (b_z-b_{z0})=-f_1|_{b_y=0}+ V_S
(b_z-b_{z0})
\end{equation}

From Eq.~(\ref{flux}) and Eq.~(\ref{lambda_SL}), a simple relation
between the slow characteristic and the flux function is
\begin{equation}
\lambda_{SL}|_{b_y=0}=\frac{\partial f_1}{\partial b_z}\Big | _{b_y=0},
\end{equation}
It is then easy to show that,
\begin{equation}
\int_{b_{z0}}^{b_z} (\lambda_{SL}|_{b_y=0}-V_S) db_z= -F_z|_{b_y=0}
\end{equation}

This indicates that a stationary point $b_z$, where $F_z=0$, will be
located where the integral on the LHS is zero. This is called the
equal-area rule. From this relation, with a given $b_{z0}$ and $b_z$,
we can determine the shock speed $V_S$ that causes the integral to
vanish. Or for a given $b_{z0}$ and $V_S$, we can determine the
possible downstream state $b_z$. We apply it to the following examples
to demonstrate the formation of intermediate shocks (which have a
super-intermediate to sub-intermediate transition) in isotropic MHD.

\begin{figure}
\includegraphics[width=15cm]{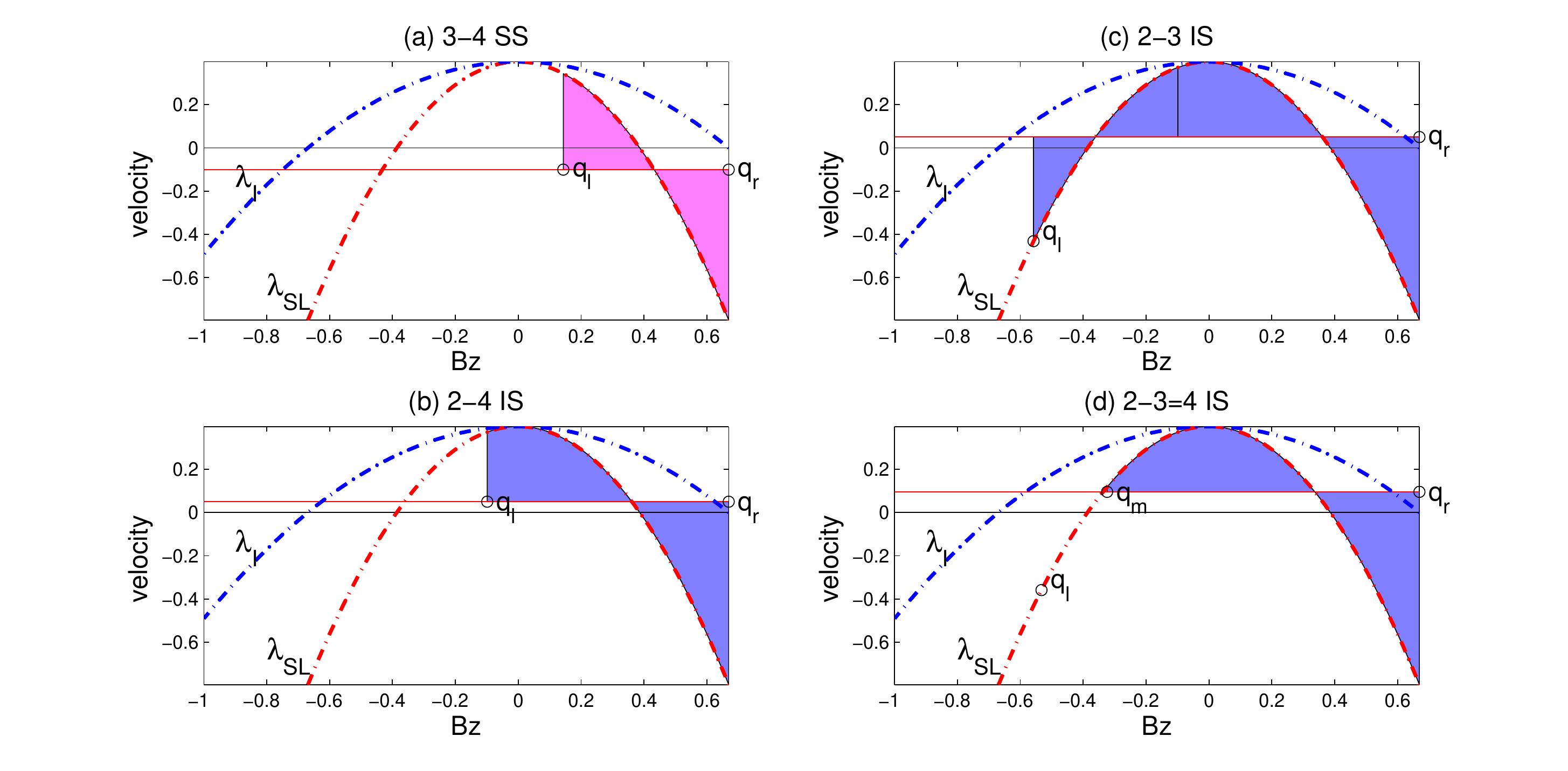} 
\caption{Application of equal-area rules with cases in ideal
(isotropic) MHD. The $\lambda_I$ and $\lambda_{SL}$ along $b_z$ are
measured in the upstream intermediate frame with upstream (point ${\bf
q}_r$) parameters, $\theta_0=42^\circ, \beta_0=1,
\varepsilon_0=1$. The vertical axis measures speed (normalized to
$C_{An}$). Once the ${\bf q}_l$ is chosen, the shock speed $V_S$
(measured by the red horizontal line) can be determined by equating
area (between the $\lambda_{SL}$ and the red line) above the red line
to area below the red line. This rule results in shock speeds (a)
$V_S= -0.1$; (b) $V_S= 0.05$; (c) $V_S= 0.05$; (d) $V_S= 0.0958$.}
\label{SS2IS}
\end{figure}

When the upstream ${\bf q}_r=(b_{z0},0)$ is given and fixed, we can
vary ${\bf q}_l=(b_z,0)$ to see the effect on possible shock
solutions. In Fig.~\ref{SS2IS}(a), when ${\bf q}_l$ is chosen above
$b_z=0$, a slow shock solution is found by determining a proper
horizontal line (${\bf q}_r-{\bf q}_l$; note that the vertical
position measures the shock speed $V_S$), which makes the red area
below the line ${\bf q}_r-{\bf q}_l$ equal the red area above. The
shock speed is slower than the upstream intermediate speed (black
horizontal line across 0), the upstream (point ${\bf q}_r$) is
super-slow and sub-intermediate ($V_S > \lambda_{SL}|_{{\bf q}_r}$,
$V_S < \lambda_{I}|_{{\bf q}_r}$. Since $\lambda$ equivalents to
$C-u$ where $C$ is the phase speed and $u$ is the bulk flow speed
measured in upstream intermediate frame, $V_S
\gtrless \lambda$ implies that the mach number measured in shock frame
$M \equiv(V_S+u)/C\gtrless 1$) and the downstream (point ${\bf q}_l$)
is sub-slow ($V_S <\lambda_{SL}|_{{\bf q}_l}$). Traditionally in
isotropic MHD, the super-fast state is termed number 1, sub-fast and
super-intermediate is 2, sub-intermediate and super-slow is 3, and
sub-slow is 4. Therefore a slow shock is also called a 3-4 SS.

In Fig.~\ref{SS2IS} (b), if ${\bf q}_l$ is chosen below the point
$b_z=0$, a 2-4 intermediate shock (${\bf q}_r-{\bf q}_l$) is formed,
with upstream being super-intermediate and downstream being
sub-intermediate and sub-slow. In Fig.~\ref{SS2IS}(c), with the same
shock speed, a 2-3 IS transitions to a ${\bf q}_l$ with a more
negative value is also possible. Note that the jump cross a compound
2-3 IS/ 3-4 SS (from this ${\bf q}_l$ to the ${\bf q}_l$ in (b) )
equals to that of the 2-4 IS in (b). In Fig.~\ref{SS2IS}(d), with the
same ${\bf q}_l$ of Fig.~\ref{SS2IS}(c), a 2-3=4 IS (${\bf q}_r-{\bf
q}_m$) with the maximum IS speed could be formed and attached by a
slow rarefaction (${\bf q}_m-{\bf q}_l$). This is a compound IS/SR
wave, with $b_z|_{{\bf q}_m}=-b_{z0}/2$ which can also be determined
by $\lambda_{SL}(b_z|_{{\bf q}_m})=[f_1(b_{z0})-f_1(b_z|_{{\bf
q}_m})]/(b_{z0}-b_z|_{{\bf q}_m})$, as shown by Brio and Wu
\cite{brio88a}. Similar arguments can be made in a system with fast
and intermediate modes.

Therefore, an intermediate shock is not directly associated with an
intermediate mode. It is steepened by magneto-sonic waves (slow or
fast modes), not by intermediate mode itself. This was first justified
by Wu's (1987) coplanar simulations \cite{wu87a} ({\it i.e.,} no
out-of-plane magnetic field is allowed), where the intermediate shock
forms even though the intermediate mode is not included (since the
out-of-plane $\delta B_y$ is necessary for nontrivial solutions of the
intermediate mode, as shown in Appendix II). The coupling of
intermediate and magneto-sonic waves and the admissibility of
intermediate shocks in the ideal MHD system was discussed by Kennel
{\it et al.} \cite{kennel90a}.

%\bibliography{paper}

\newpage

\end{document}